\documentclass[%
 reprint,
superscriptaddress,
showpacs,preprintnumbers,
 amsmath,amssymb,
 aps,
floatfix,
]{revtex4-1}

\usepackage{mathtools}
\usepackage{graphicx}
\usepackage{dcolumn}
\usepackage{bm}
\usepackage{hhline}
\usepackage{color,soul}
\usepackage{float}
\usepackage[normalem]{ulem}
\usepackage[dvipsnames]{xcolor}
\usepackage{hyperref}
\hypersetup{
    colorlinks,
    linkcolor={red!80!black},
    citecolor={blue!80!black},
    urlcolor={blue!80!black}
}

\newcommand{\gf}{\mathbf{G}(\mathbf{q})}
\newcommand{\gfel}[2]{G_{#1}^{#2}(\mathbf{q})}

\begin{document}


\title{Exploring the link between crystal defects and non-affine displacement fluctuations}

\author{Pankaj Popli}
\affiliation{%
Tata Institute for Fundamental Research,  Centre for Interdisciplinary Sciences, 36/P Gopanapally, Hyderabad 500107,  India.
}%
\author{Sayantani Kayal}%
\affiliation{%
 Tata Institute for Fundamental Research,  Centre for Interdisciplinary Sciences, 36/P Gopanapally, Hyderabad 500107,  India.
}%

\author{Peter Sollich}
\affiliation{%
 King's College London, Department of Mathematics, Strand, London WC2R 2LS, UK}
 \affiliation{Institute for Theoretical Physics,  University of G\"ottingen, 37077 G\"ottingen, Germany 
}%
\author{Surajit Sengupta}
\affiliation{%
 Tata Institute for Fundamental Research,  Centre for Interdisciplinary Sciences, 36/P Gopanapally, Hyderabad 500107,  India.
}%

\date{\today}
\begin{abstract}
We generalize, and then use, a recently introduced formalism to study thermal fluctuations of atomic displacements in several two and three dimensional crystals. We study both close packed as well as open crystals with multi atom bases. 
Atomic displacement fluctuations in a solid, once coarse-grained over some neighborhood may be decomposed into two mutually orthogonal components. In any dimension $d$ there are always $d^2$ {\it affine} displacements representing local strains and rotations of the ideal reference configuration. In addition, there exists a number of {\it non-affine} localized displacement modes that cannot be represented as strains or rotations. The number of these modes depends on $d$ and the size of the coarse graining region. All thermodynamic averages and correlation functions concerning the affine and non-affine displacements may be computed within a harmonic theory. We show that for compact crystals, such as the square and triangular in $d=2$ and the simple, body-centered and face-centered cubic crystals in $d=3$, a single set of $d-$fold degenerate modes always dominate the non-affine sub-space and are separated from the rest by a large gap. These modes may be identified with specific precursor configurations that lead to lattice defects. In open crystals, such as the honeycomb and kagome lattices, there is no prominent gap although soft non-affine modes continue to be associated with known floppy modes representing localized defects. Higher order coupling between affine and non-affine components of the displacements quantify the tendency of the lattice to be destroyed by large homogeneous strains. We show that this coupling is larger by almost an order of magnitude for open lattices as compared to compact ones. Deformation mechanisms such as lattice slips and stacking faults in close packed crystals can also be understood within this framework. The qualitative features of these conclusions are expected to be independent of the details of the atomic interactions.
 
\end{abstract}

\maketitle


\section{\label{sec:level1}Introduction}
A crystalline solid is formed from the liquid when translation and orientation symmetries are broken as a result of a thermodynamic phase transformation~\cite{CL}. At any non-zero temperature, atomic fluctuations in the crystalline solid tend to restore those symmetries. The current paradigm classifies atomic fluctuations as being either ``smooth" or ``singular". The former comprise long wavelength, hydrodynamic, smooth variations of the elastic displacements and density fields~\cite{born,martin}, while the latter are defects, where the displacement becomes discontinuous~\cite{rob,hirth,nabarro}.  Such a classification has proved to be immensely useful in understanding many commonly observed properties of solids~\cite{ashcroft} as well as melting~\cite{CL} and failure of solids in response to external mechanical loads~\cite{rob,seth-review}. This viewpoint has also been extended with some success towards understanding the mechanical properties of amorphous solids~\cite{falk-review}.  However, in contrast to crystals, the lack of long ranged structural order in amorphous solids precludes a clear distinction between smooth and singular displacements~\cite{wijtman,acharya}. It has been suggested (and experimentally observed) that localized, {\it non-singular} displacement configurations  play the r\^ole of lattice defects in such solids~\cite{falk-review, chik,schall}.   

For a while now, there has been an effort by some of us to formulate a different way of classifying atomic displacements in solids, which we hope, may have more general applicability~\cite{sas1,sas2,sas3,sas4,sas5,sas6,sas7,poplitrap,pnas,protein}. Borrowing an idea first used to study mechanical deformation in glasses~\cite{falk}, it was shown that any set of atomic displacements of an atom and its neighbors, within a specified ``coarse graining" region,  may be decomposed into two mutually orthogonal sub-spaces using a well defined projection formalism~\cite{sas1,sas2,sas3}. The {\it affine} component of these displacements represents homogeneous linear transformations of a reference configuration within the coarse graining volume. Ignoring trivial uniform displacements, these are isotropic expansion, shear strains and local rotations. Since, in general, not all displacements can be described completely by these linear transformations, inevitably, a {\it non-affine} component remains. By construction, the affine and the non-affine parts of the displacements are linearly independent~\cite{sas1}. Thermodynamically conjugate fields may be defined, which enhance or suppress each part independently of the other. The affine displacements couple to local stresses (and torques) while the non-affine component of the displacement couples to a new ``non-affine'' field~\cite{sas2,pnas,poplitrap}. Enhancing non-affine fluctuations by increasing temperature, applying large strains or the non-affine field leads to the creation of defects~\cite{sas2,sas5,pnas}. Indeed, atomic fluctuations that act as precursors to the formation of defects such as dislocation dipoles have been shown to be the most prevalent, though not the sole, non-affine displacement even within a small oscillation, harmonic, approximation~\cite{sas2,sas5}.  

This decomposition of displacements into affine and non-affine have lead to a deeper understanding in several disparate contexts. 
For example, it has been possible to show that rigid solids are always thermodynamically metastable for infinitesimal stresses. Loading a rigid solid is tantamount to quenching it across an equilibrium first order phase transition~\cite{pnas}. Yielding of a crystal under load is simply the decay of this metastable phase to the stable phase, which eliminates stress by non-affine atomic rearrangements. This radically different viewpoint nevertheless allows one to calculate, for the first time, strain rate dependent yielding thresholds using classical nucleation theory. 

In networked solids, where atoms are bound by strong chemical bonds, dislocations do not form. Nevertheless such solids may deform by special singular formations called pleats or ``ripplocations", which have been described within the same language of non-affine displacements~\cite{sas4,sas7}. 

It has further been shown that colloidal crystals with any given interaction may be arranged in any structure whatsoever if it is only possible to suppress non-affine displacements away from this reference configuration. Special, dynamic, feedback controlled laser traps have been proposed, though not yet experimentally realized, which may be able to perform this feat.  Unlike static traps, the structures stabilized by such a process are translationally invariant and possess all allowed zero modes~\cite{poplitrap}. 

Finally, in a {\it protein} -  a large molecule consisting of interacting atoms with no spatial long ranged order - it has been shown that important conformational changes, which precede binding to ligands are always non-affine. These may be discovered by simply projecting out the local atomic displacements using the same projection formalism. Regions with large susceptibility for non-affine displacements correlate with binding hotspots and spatial correlations of the magnitude of non-affine ness mark sites of allosteric control~\cite{protein}. 

In this paper, we carry this program forward by returning to a study of small amplitude non-affine displacements in periodic crystals. We attempt to answer some questions that naturally arise from our previous work. So far, except for the work on proteins~\cite{protein}, the non-affine analysis has been mainly restricted to two dimensional solids for simplicity. Most experiments, however, are in three dimensions and an extension of our work to three dimensions is therefore necessary. We have also felt the need for a comparative  study of harmonic non-affine fluctuations among different kinds of crystals, especially because of their importance as precursors to lattice defects. Finally, we would like to test the robustness of our results to variation of some parameters such as the extent of coarse graining and nature of interactions.  One of the aims of the present work is to discover features that are common to all crystalline solids and differentiate them from those that depend on details of the crystal structure, dimensionality and interaction parameters. 

Our main results are as follows. The eigenvalue spectrum of the Hessian of the coarse-grained Hamiltonian taken with respect to local atomic displacements and projected onto the non-affine subspace~\cite{sas2}, always shows a prominent {\it gap} between the largest eigenvalues and the others, for all Bravais lattices with a monatomic basis. The gap increases with the size of the coarse graining volume. The dominant non-affine eigenmode corresponding to the largest eigenvalue features displacements that may be identified with defect precursors~\cite{sas2,sas3}. For open lattices featuring a multi atom basis, the gap is much less prominent, although large eigenvalue floppy modes continue to resemble precursors for known defects. The relative prominence of modes in the non-affine eigenvalue spectrum for open lattices is more sensitive to the nature of the interactions compared to those in close packed crystals. Spatial correlation functions of the affine and non-affine modes are similar in nature among the various crystal structures studied. Affine and non-affine modes couple at, higher than linear, order~\cite{sas1}. This coupling measures the susceptibility of the crystal to producing non-affine displacements in response to small external stress. We find that open crystals are more susceptible than close packed ones by almost an order of magnitude.

The rest of the paper is arranged as follows. In the next section (Section~\ref{theory}) we introduce the projection formalism for atomic displacements. Some of this material has been discussed in our earlier publications but we generalize the derivation here to make it directly applicable to crystals with a multi-atom basis and in any dimension. This is followed by Section~\ref{models} where we present a brief description of the crystal lattices studied and the interaction models. In two dimensions we study the triangular, square, honeycomb and kagome lattices, while in three dimensions we restrict ourselves to cubic crystals and study the simple cubic, body centered cubic and face centered cubic structures. In Section ~\ref{results} we present our results for several quantities related to the statistics of displacement fluctuations, critically comparing them among the different lattices. In Section~\ref{discuss} we summarize our main findings in detail by discussing our main results and analyzing their implications, especially with respect to indications to directions of future research.

\section{\label{theory}Non affine fluctuations: The projection method}
In this section we present the method we use to project out local atomic displacements into affine and non-affine components, extending it from earlier formulations~\cite{sas1,sas2,sas3} by 
taking care that the derivation is directly applicable to crystals with a multi-atom basis in any dimension.  
In the most general setting, non-affine displacements are those atomic displacements that cannot be captured by a homogeneous deformation. For example, imagine an ideal (defect free) crystal at zero temperature consisting of atoms placed at their reference position. The Cauchy-Born rule~\cite{cbr} (CBR) states that any external deformation caused by changing the shape of the boundary of the solid is distributed homogeneously among the atoms of the crystal, which are shifted appropriately from their reference position. This, of course, amounts to stating that for an ideal crystal there are no non-affine deformations at zero temperature.  At finite temperature, one expects this rule to hold on the average, with the 
(finite size scaled) elastic moduli~\cite{zahn,kerst1} setting the scale for the displacement fluctuations averaged within some local coarse graining volume $\Omega$. While this is more or less true, there is a subtle point here that needs attention. Indeed,
%
%
%
atomic displacement fluctuations within $\Omega$ decompose naturally into {\em two} mutually orthogonal subspaces. One, designated as affine, consists of fluctuations for which the CBR holds locally and instantaneously, while the other is the set of non-affine displacement modes for which the CBR is violated. While the former may be directly connected to the elastic moduli, the latter represents fluctuations that act as precursors for defects. We show below how this decomposition may be carried out for a generic crystalline solid.  
%
%
%
%
%
We consider a $d$ dimensional lattice with $N$ lattice sites and $ N_b$ basis particles per site. The total number of particles in the system is $N\times N_b$. We take $\{\mathbf{R}_{i\alpha}\}$ as the equilibrium position vector for any site $i  \in \{0,1,2 \dots N-1\}$, where $\alpha \in \{0,1,2 \dots N_b-1\}$ represents the index of a basis particle. To distinguish affine and non-affine displacements we consider relative displacements of pairs of atoms whose reference positions are within some fixed coarse-graining distance $\mathfrak{r}$ of each other. Specifically, the coarse-graining region around the basis on lattice site $i$ is defined as
\begin{equation}
\Omega(i) = \{(j\gamma,i\alpha)| 0<|\mathbf{R}_{j\gamma}-\mathbf{R}_{i\alpha}| \leq \mathfrak{r} \land (j\neq i \lor \gamma>\alpha)\}.
\label{Omega_def}
\end{equation}
In words, $\Omega(i)$ contains all pairs of indices  $(j\gamma, i\alpha)$ of particles within the specified dsitance $\mathfrak{r}$; at least one of these particles has to belong to the basis around lattice site $i$. The last constraint in Eq.(\ref{Omega_def}) merely avoids double-counting of pairs {\em within} the basis, by insisting that index pairs of the form $(i\gamma,i\alpha)$ have got ordered basis indices, $\gamma < \alpha$. We denote the number of particle pairs in $\Omega(i)$ by $N_{\Omega}$ and number the elements of $\Omega(i)$ in some arbitrary fashion as\[
\Omega(i) = \{ (j_n\gamma_n, i\alpha_n), n = 1\ldots N_{\Omega}  \}
\]
%
%
%
%
%
We note that the lattice symmetries mean that all neighbourhoods $\Omega(i)$ for different $i$ are just translated copies of each other.

When the lattice is deformed particles undergo displacements and take new positions $\{\mathbf{r}_{i\alpha}\}$; we write the displacement from their equilibrium positions as $\mathbf{u}_{i\alpha} = \mathbf{r}_{i\alpha} - \mathbf{R}_{i\alpha}$. It has been shown\cite{sas1,sas2,sas3} that the displacements in a  given deformed coarse-graining volume can be expressed as a linear combination of independent affine and non-affine deformations. For a fully affine deformation of the coarse-graining volume around lattice site $i$ there is by definition a local $d\times d$ dimensional deformation matrix $\mathcal{D}$  such that
\begin{equation}
\mathbf{u}_{j_n\gamma_n,i\alpha_n} = \mathcal{D}\mathbf{R}_{j_n\gamma_n,i\alpha_n}, \qquad  n = 1 \ldots N_{\Omega}
\end{equation}
using the abbreviations 
 $\mathbf{u}_{j_n\gamma_n,i\alpha_n} = \mathbf{u}_{j_n\gamma_n}-\mathbf{u}_{i\alpha_n}$ and $\mathbf{R}_{j_n\gamma_n,i\alpha_n} = \mathbf{R}_{j_n\gamma_n}-\mathbf{R}_{i\alpha_n}$. 
 In general, $\{\mathbf{u}_{i\alpha}\}$ will have contributions from non-affine transformations as well. In such cases $\mathcal{D}$ is defined as the matrix that minimizes 
 %
 %
%
%
\begin{equation}\label{eq:def_chi}
\chi_i=\min_{\mathcal{D}}\left(\sum_{n=1}^{N_{\Omega}}\left(\mathbf{u}_{j_n\gamma_n,i\alpha_n}-\mathcal{D}\mathbf{R}_{j_n\gamma_n,
i\alpha_n}\right)^2\right)
\end{equation}
Therefore $\chi_i$ is a measure of the {\em non-affinity} at  the given lattice site $i$. 
%
%
%
%

We now introduce some simplified notation  by rearranging components of $\mathbf{u}_{j_n\gamma_n}-\mathbf{u}_{i\alpha_n}$ for all $n$ into a column vector $\mathbf{\Delta}$ of length $N_{\Omega}d$ with elements
\[
\Delta_{n\mu} = {u}_{j_n\gamma_n,i\alpha_n}^{\mu}
\]
where $\mu=1\ldots d$ denotes the spatial components of the displacement vectors.
Similary a column vector $\mathbf{e}$ of length $d^2$ is obtaind by arranging the elements of $\mathcal{D}$ in order $(\mathcal{D}_{11},\mathcal{D}_{12}...\mathcal{D}_{1d}...\mathcal{D}_{d1},\mathcal{D}_{d2}...\mathcal{D}_{dd})$ . With these definitions, Eq.~(\ref{eq:def_chi}) may be written concisely as 
\begin{equation}\label{eq:def_chi2}
\chi=\min_{\mathbf{e}}\left( \left[\mathbf{\Delta}-\mathsf{R}\mathbf{e}\right]^2\right)
\end{equation}
Here we have introduced a block matrix $\mathsf{R}$ 
whose elements are given by
\[
\mathsf{R}_{n\mu,\nu\nu'}=\delta_{\mu\nu}R_{j_n\gamma_n,i\alpha_n}^{\nu'}
\]
and the entries of the vector $\mathsf{R}\mathbf{e}$ are given by $\sum_{\nu\nu'}\mathsf{R}_{n\mu,\nu\nu'}e_{\nu\nu'}$ in the obvious way.
Once Eq.~(\ref{eq:def_chi2}) is minimized, we  obtain the residual contribution from non-affine deformation and the ``best-fit" affine strain. These can be expressed as
\begin{eqnarray}
\cr\chi &=& \mathbf{\Delta}^{\rm{T}}\mathsf{P}\mathbf{\Delta} \label{eq:chi}\\
\cr\mathbf{e} &=&\mathsf{Q}\mathbf{\Delta}.
\label{eq:strain}
\end{eqnarray}
 with  the matrices $\mathsf{P} = \mathsf{I}-\mathsf{RQ}$ and $\mathsf{Q}=\left(\mathsf{R}^{\rm{T}}\mathsf{R}\right)^{-1}\mathsf{R}^{\rm{T}}$. Note that $\mathsf{P}$  is a projection matrix ($\mathsf{P}^2 = \mathsf{P}$), which, when acting on the space of $\mathbf{\Delta}$, extracts only the non-affine component of the displacements. It can be seen that $\mathsf{P}$ has $d^2$ zero and $N_{\Omega}d-d^2$ non-zero eigenvalues corresponding to the affine and non-affine subspaces respectively. As usual $\mathsf{I-P}$ will then project out the affine component of $\mathbf{\Delta}$. The elements of the best-fit affine transformation matrix $\mathcal{D}$ from Eq.~(\ref{eq:strain}) can be written explicitly as 
 \begin{equation}\label{eq:fit_e_first}
\mathcal{D}_{\mu\nu}=
\sum_{\nu'} ({\bf M}^{-1})_{\nu\nu'} 
\sum_{n}
R_{j_n\gamma_n,i\alpha_n}^{\nu'} \Delta_{n\mu}
\end{equation}
in terms of the matrix ${\bf M}$ with elements
\[
M_{\nu\nu'} = \sum_n
R_{j_n\gamma_n,i\alpha_n}^\nu
R_{j_n\gamma_n,i\alpha_n}^{\nu'}
\]
For the lattices considered in this paper, $\bf{M}$ is diagonal due to lattice symmetries (see also the discussion in Ref.~\cite{sas1}) so that Eq.(\ref{eq:fit_e_first}) simplifies to
\begin{equation}\label{eq:fit_e}
\mathcal{D}_{\mu\nu}=\frac{1}{\sum_{n}
(R_{j_n\gamma_n,i\alpha_n}^{\nu})^2}\sum_{n}
R_{j_n\gamma_n,i\alpha_n}^{\nu} \Delta_{n\mu}
\end{equation}


We now obtain the statistics of $(\chi , \mathbf{e})$ in the classical canonical ensemble for any lattice and dimension. For any given Hamiltonian $\mathcal{H}$ the canonical probability distribution is  
\begin{equation}
\mathbb{P}(\mathbf{p},\mathbf{u})= \frac{1}{Z}e^{-\beta \mathcal{H}\left( \mathbf{p},\mathbf{u}\right)}
\end{equation}
Here we restrict the Hamiltonian to the {\em harmonic approximation}, 
\begin{equation}
\mathcal{H}=\sum_{i\alpha}\frac{\mathbf{p}_{i\alpha}^2}{2m_{i\alpha}}+\frac{1}{2}\sum_{ i\alpha,j\gamma}\sum_{\mu\nu}u_{i\alpha}^{\mu}\phi_{i\alpha,j\gamma}^{\mu\nu}u_{j\gamma}^{\nu}
\end{equation}
where ${\bf p}_{i\alpha}$ are the momenta, $m_{i\alpha}$ the masses and $\phi_{i\alpha,j\gamma}^{\mu\nu}$ are the elements of the Hessian. The Hamiltonian can be easily diagonalized if one takes a plane wave ansatz for the displacements. We therefore write 
\begin{equation}
\mathbf{u}_{i\alpha} = \int  \frac{d\mathbf{q}}{V_{BZ}}\mathbf{u}_{\alpha}(\mathbf{q})e^{i\mathbf{q}\cdot {\bf R}_{i\alpha} } \nonumber \\
\end{equation}
and similarly
\begin{equation}
\mathbf{u}_{i\alpha,j\gamma} = \int\frac{d\mathbf{q}}{V_{BZ}} \mathbf{u}_{\alpha}(\mathbf{q})\left(e^{i\mathbf{q}\cdot \mathbf{R}_{i\alpha}} - e^{i\mathbf{q}\cdot \mathbf{R}_{j\gamma}}\right)\nonumber
\end{equation}
%
where the integration runs over the first Brillouine zone with volume $V_{BZ}$ and $\mathbf{q}$ is the wave vector. The 
Lattice Green's Function (LGF) may be obtained as the inverse $\gf=\mathbf{D}^{-1}(\mathbf{q})$ of the dynamical matrix $\mathbf{D}(\mathbf{q})$ with elements
\begin{equation}
D_{\alpha\gamma}^{\mu\nu}(\mathbf{q})=\sum_{i}\phi_{i\alpha,0\gamma}^{\mu\nu} e^{i\mathbf{q}\cdot\left(\mathbf{R}_{i\alpha}-\mathbf{R}_{0\gamma}\right)}
\end{equation}
With the knowledge of the LGF, thermal averages of different quantities are easy to calculate. For example the displacement correlator reads in Fourier space
$$\left\langle u_{\alpha}^{\mu}(\mathbf{q})u_{\gamma}^{\nu}(\mathbf{q}')\right\rangle=\gfel{\alpha\gamma}{\mu\nu}\delta(\mathbf{q}+\mathbf{q}')V_{BZ},$$  and in real space 
\begin{eqnarray}
\left\langle u_{i\alpha}^{\mu} u_{j\gamma}^{\nu}\right\rangle=\dfrac{1}{\beta}\int \frac{d\mathbf{q}}{V_{BZ}}\,\gfel{\alpha\gamma}{\mu\nu} e^{i \mathbf{q}\cdot \left(\mathbf{R}_{i\alpha}-\mathbf{R}_{j\gamma}\right)}.
\label{corr_real_space}
\end{eqnarray}
Along similar lines, for our coarse-graining volume one can obtain the thermal average of any observable $A(\mathbf{\Delta})$ as
\begin{equation}\label{eq:avrg_on_omega}
\left\langle A(\mathbf{\Delta})\right\rangle=\frac{1}{\mathcal{Z}}\int d\mathbf{\Delta}\, A(\mathbf{\Delta})e^{-\frac{1}{2}\mathbf{\Delta}^{\rm{T}}\mathsf{C}^{-1}\mathbf{\Delta}} 
\end{equation}
with the normalisation constant   $\mathcal{Z} = (2\pi)^{N_{\Omega}d/2}\sqrt{|\hspace{2pt}\mathsf{C}\hspace{2pt}|}$.
The covariance matrix $\mathsf{C}$ in the above equation can be obtained from the LGF via 
\begin{eqnarray}
\mathsf{C}_{n\mu,m\nu} &=& \left\langle\Delta_{n\mu}\Delta_{m\nu}\right\rangle 
\nonumber\\
&=&\int \dfrac{d\mathbf{q}}{\beta V_{BZ}} 
\bigg[
\gfel{\gamma_n\gamma_m}{\mu\nu}
e^{i \mathbf{q}\cdot\mathbf{R}_{j_n\gamma_n,j_m\gamma_m}}\label{eq:cmat}
\\&&{}
-\gfel{\alpha_n\gamma_m}{\mu\nu}
e^{i \mathbf{q}\cdot\mathbf{R}_{i\alpha_n,j_m\gamma_m}}
\nonumber\\&&{}
-\gfel{\gamma_n\alpha_m}{\mu\nu}
e^{i \mathbf{q}\cdot\mathbf{R}_{j_n\gamma_n,i\alpha_m}}
\nonumber\\&&{}
+ 
\gfel{\alpha_n\alpha_m}{\mu\nu}
e^{i \mathbf{q}\cdot\mathbf{R}_{i\alpha_n,i\alpha_m}}
\bigg]\nonumber
\end{eqnarray}
To obtain the statistics of $(\chi,\mathbf{e})$ we make use of Eq.~(\ref{eq:avrg_on_omega}) and obtain the characteristic function $\rm{\Phi}\left(\rm{k},\pmb{\kappa}\right)$ for the joint probability distribution $\mathbb{P}(\chi,\mathbf{e})$. 
\begin{eqnarray}\label{eq:chrtr_fun}
\rm{\Phi}\left(\rm{k},\pmb{\kappa}\right)&=&
\rm{exp}\left({-\frac{1}{2}\pmb{\kappa}^{\rm{T}}\mathsf{QC}\left(\mathsf{I}-2ik\mathsf{PC}\right)^{-1}\mathsf{Q}^{\rm{T}}\pmb{\kappa}}\right)\\\nonumber
&&\times\frac{1}{\sqrt{|\hspace{2pt}\mathsf{I}-2ik\mathsf{PCP}\hspace{2pt}|}}.
\end{eqnarray}
Using the identity $$\left(\mathsf{I}-2ik\mathsf{PC}\right)^{-1}=\mathsf{I}+\left(\mathsf{I}-2ik\mathsf{PC}\right)^{-1}\left(2ik\mathsf{PC}\right)$$ the last result can be written in terms of the characteristic function for the marginals as follows:
\begin{equation}
\rm{\Phi}\left(\rm{k},\pmb{\kappa}\right)=\mathit{\Phi}_{\chi}(k)\mathit{\Phi}_{\mathbf{e}}(\pmb{\kappa})e^{-ik\pmb{\kappa}^{\rm{T}}\mathsf{QC}\left(\mathsf{I}-2ik\mathsf{PC}\right)^{-1}\mathsf{PC}\mathsf{Q}^{\rm{T}}\pmb{\kappa}}
\end{equation}
where
\begin{eqnarray}
\mathit{\Phi}_{\chi}(k)&=&\frac{1}{\prod\limits_l \sqrt{1-2ik\sigma_l}}\\
\mathit{\Phi}_{\mathbf{e}}(\pmb{\kappa})&=&e^{-\frac{1}{2}\pmb{\kappa}^{\rm{T}}\mathsf{QC}\mathsf{Q}^{\rm{T}}\pmb{\kappa}}
\end{eqnarray}
and the $\sigma_l$ are the eigenvalues of $\mathsf{PCP}$. 
For $\pmb{\kappa} = 0$ and $k=0$, $\rm{\Phi}\left(\rm{k},\pmb{\kappa}\right)$ reduces to $\mathit{\Phi}_{\chi}(k)$ and $\mathit{\Phi}_{\mathbf{e}}(\pmb{\kappa})$ respectively.  The term in the exponential governs the (non-linear) coupling between the non-affine and affine components of the displacements. Previous work has shown that this coupling is significant only for large uniaxial and shear strains~\cite{sas1}. For smaller strains, it can largely be ignored. 

With the knowledge of the  characteristic function, thermal averages and other higher order moments may be computed such as
\begin{equation}
\left\langle\chi\right\rangle=\rm{Tr}\left(\mathsf{PCP}\right)
\label{trace}
\end{equation}
\begin{equation}
\left\langle\mathbf{e}\mathbf{e}^{\rm{T}}\right\rangle=\mathsf{QC}\mathsf{Q}^{\rm{T}}
\end{equation}
From Eq.~(\ref{trace}) it is clear that $\langle\chi\rangle$ is a sum over the eigenvalues of  $\left(\mathsf{PCP}\right)$. Each eigenvalue represents the contribution of a specific {\em non-affine mode} to $\chi$. It has been shown~\cite{sas2} that these eigenvalues are elements of the inverse Hessian of the free energy in the direction of the non-affine mode in configuration space. A large eigenvalue implies a small value of the local curvature of the free energy minimum. We shall see later in this paper that the corresponding {\it eigenvectors} are precisely those atomic displacement fluctuations that lead to lattice defects or other imperfections tending to destroy crystalline order. The non-affine mode corresponding to the largest eigenvalue therefore has the highest contribution to this process.

The non-affine parameter $\chi$ depends linearly on temperature $T$ and is inversely proportional to the strength of the interaction. Due to the fact that the underlying distribution of $\mathbf{\Delta}$ is Gaussian, $\left\langle \mathbf{e}\right\rangle$ vanishes unless an external stress is present. Finally, the leading order non-linear coupling between non-affine and affine modes is given by
\begin{equation}
\left\langle\chi\mathbf{e}\mathbf{e}^{\rm{T}}\right\rangle-\left\langle\chi\right\rangle\left\langle\mathbf{e}\mathbf{e}^{\rm{T}}\right\rangle=2\mathsf{QC[P,C]}\mathsf{Q}^{\rm{T}}.
\end{equation}

Two-point distributions and spatial correlation functions for $\chi$ and  $\mathbf{e}$ may also be calculated following the procedure explained in Refs.~\cite{sas1,sas2,sas3}. Below we include a brief description for completeness.  

The spatial correlations of the non-affine parameter are given by 
\begin{equation}
\langle\chi\bar{\chi}\rangle - \langle\chi\rangle\langle\bar{\chi}\rangle = 2\mathsf{Tr(P\bar{C}P)(P\bar{C}P)^T}
\end{equation} 
The two point covariance $\mathsf{\bar{C}}$ between relative displacements within two coarse-graining neighbourhoods $\Omega\equiv\Omega(i)$ and $\bar\Omega\equiv\Omega(k)$ around lattice positions $\mathbf{R}_{i0}$ and $\mathbf{R}_{k0}$, respectively, is defined as 
\begin{equation}
\bar{\mathsf{C}}_{n\mu,m\nu}
= \langle\Delta_{n\mu}\bar\Delta_{m\nu}\rangle
\end{equation}
It is obtained from an expression similar to Eq.~(\ref{eq:cmat}) (see Ref.~\cite{sas1} for details):
\begin{eqnarray}
\bar{\mathsf{C}}_{n\mu,m\nu} 
&=&\int \dfrac{d\mathbf{q}}{\beta V_{BZ}} 
\bigg[
\gfel{\gamma_n\gamma_m}{\mu\nu}
e^{i \mathbf{q}\cdot\mathbf{R}_{j_n\gamma_n,l_m\gamma_m}}\label{eq:cmat2pnt}
\\&&{}
-\gfel{\alpha_n\gamma_m}{\mu\nu}
e^{i \mathbf{q}\cdot\mathbf{R}_{i\alpha_n,l_m\gamma_m}}
\nonumber\\&&{}
-\gfel{\gamma_n\alpha_m}{\mu\nu}
e^{i \mathbf{q}\cdot\mathbf{R}_{j_n\gamma_n,k\alpha_m}}
\nonumber\\&&{}
+ 
\gfel{\alpha_n\alpha_m}{\mu\nu}
e^{i \mathbf{q}\cdot\mathbf{R}_{i\alpha_n,k\alpha_m}}
\bigg]\nonumber
\end{eqnarray}
%
where we have assumed that the elements of $\Omega(k)$ are numbered $(l_m\gamma_m,k\alpha_m)$.
For all simple lattices in $2d$ and $3d$ with one particle basis, the correlations $\langle\chi\bar{\chi}\rangle - \langle\chi\rangle\langle\bar{\chi}\rangle$ are short ranged. 

Strain-strain correlation may be obtained from
\begin{equation}
\left\langle\mathbf{e}\bar{\mathbf{e}}^{\rm{T}}\right\rangle=\left\langle\mathsf{Q}\mathbf{\Delta}\bar{\mathbf{\Delta}}^{\rm{T}}\mathsf{Q}^{\rm{T}}\right\rangle=\mathsf{Q\bar{C}}\mathsf{Q}^{\rm{T}}.
\end{equation} 
It is often useful to consider these correlations in Fourier space, where they can be expressed as~\cite{sas1}
\begin{equation}
\left\langle\mathbf{e}\bar{\mathbf{e}}^{\rm{T}}\right\rangle(\mathbf{q})=\mathsf{Q\bar{C}}(\mathbf{q})\mathsf{Q}^{\rm{T}}
\end{equation} with the Fourier transform $\mathsf{\bar{C}}(\mathbf{q})$ defined via
$$\bar{\mathsf{C}}_{n\mu,m\nu} = \int \frac{d\mathbf{q}}{V_{BZ}} \bar{\mathsf{C}}_{n\mu,m\nu}(\mathbf{q})e^{i \mathbf{q}\cdot\mathbf{R}_{i0,k0}}.$$
Comparison with Eq.(\ref{eq:cmat2pnt}) then shows that $\beta \bar{\mathsf{C}}_{n\mu,m\nu}(\mathbf{q})$ is given directly by the term in square brackets in Eq.(\ref{eq:cmat}). This follows from the fact that the particle pairs in $\Omega(k)$ are just those in $\Omega(i)$ translated by $\mathbf{R}_{k0,i0}$; e.g.\ in the first term of Eq.(\ref{eq:cmat2pnt}) one has after extracting the Fourier factor $e^{i\mathbf{q}\cdot\mathbf{R}_{i0,k0}}$ the phase term
\begin{eqnarray*}
e^{i\mathbf{q}\cdot(\mathbf{R}_{j_n\gamma_n,l_m\gamma_m}-\mathbf{R}_{i0,k0})} &=&
e^{i\mathbf{q}\cdot[
\mathbf{R}_{j_n\gamma_n}-(\mathbf{R}_{l_m\gamma_m}-\mathbf{R}_{k0,i0})]}\\
&=&
e^{i\mathbf{q}\cdot(
\mathbf{R}_{j_n\gamma_n}-\mathbf{R}_{j_m\gamma_m})}
\\
&=& e^{i\mathbf{q}\cdot
\mathbf{R}_{j_n\gamma_n,j_m\gamma_m}}
\end{eqnarray*}
Correlations of the affine displacements viz.\ local volume change ($e_v$), uniaxial or deviatoric strain ($e_u$), shear strain ($e_s$) and rotation ($w$) respectively, may be obtained using the component forms as follow,
\begin{align}\label{eq:straincorr}
\left\langle e^{2}_{v}\right\rangle(\mathbf{q})&=E_{1111}+E_{2222}+2E_{1122}\\ \nonumber
\left\langle e^{2}_{u}\right\rangle(\mathbf{q})&=E_{1111}+E_{2222}-2E_{1122}\\ \nonumber
\left\langle e^{2}_{s}\right\rangle(\mathbf{q})&=E_{1212}+E_{2121}+2E_{1221}\\ 
\left\langle w^{2}\right\rangle(\mathbf{q})&=E_{1212}+E_{2121}-2E_{1221} \nonumber,
\end{align}
Here we have used the same notation for the fourth rank tensor, ${\mathsf E} = {\mathsf Q}\bar{\mathsf C}({\bf q}){\mathsf Q}^{\rm T}$ as in Ref.~\cite{sas1}, and $\left\langle e^{2}_{v}\right\rangle(\mathbf{q})$ etc.\ are strain correlators in Fourier space. Expressions for 3d can be obtained in similar fashion, for instance, strain correlation in 3d for volume, uniaxial and shear in the $x-y$ plane are as follows,\\\\
\begin{align}\label{eq:straincorr3d}
\left\langle e^{2}_{v}\right\rangle(\mathbf{q})&=E_{1111}+E_{2222}+E_{3333}\\ \nonumber
&+2\left(E_{1122}+E_{1133}+E_{2233}\right)\\ \nonumber \\ \nonumber
\left\langle e^{2}_{u}\right\rangle(\mathbf{q})&=E_{1111}+E_{2222}+E_{3333}\\ \nonumber 
&-2\left(E_{1122}+E_{1133}-E_{2233}\right)\\ \nonumber \\ \nonumber
\left\langle e^{2}_{s}\right\rangle(\mathbf{q})&=E_{1212}+E_{2121}+2E_{1221}.\\  \nonumber \\ \nonumber
\end{align}
Other components of the shear strain (and rotation) can be written down by analogy.  
\section{\label{models}Models}
In this paper, we use the methods of the last section to obtain the  statistics of affine and non-affine displacements for a number of lattices in two (2d) and three (3d) dimensions. In 2d we consider lattices both with a single atom basis such as the triangular and square lattices as well as those with a multi-atom basis like the planar honeycomb and the kagome lattices. In 3d we confine ourselves to a study of cubic systems, namely, the simple cubic, body centered and face cantered cubic lattices. In order to keep the discussion general we model the interactions by harmonic springs. Our results are therefore valid for any crystalline solid at sufficiently low temperatures where anharmonic effects may be neglected. A typical Hamiltonian for such interactions is
\begin{equation}
\mathcal{H}=\sum_{i\alpha}\frac{\mathbf{p}_{i\alpha}^2}{2m}+\sum_{\left\langle i\alpha,j\gamma\right\rangle}\frac{k_{i\alpha,j\gamma}}{2}\left(\mathbf{u}_{i\alpha,j\gamma}\cdot \hat{\mathbf{R}}_{i\alpha,j\gamma}\right)^2
\end{equation}

Here  $k_{i\alpha,j\gamma}$  determines the spring constant acting between particle pairs
$i\alpha,j\gamma$. 
The $k_{i\alpha,j\gamma}$ are chosen such that the lattice is stable and satisfies Maxwell's stability criteria~\cite{stability}. In particular, we take $k_{i\alpha,j\gamma}$ to be equal to $k_1$ for nearest neighbours and $k_2$ for next nearest neighbours; 
interactions beyond the second neighbor shell are neglected.
Additionally, throughout the paper, the nearest neighbour bond strength $k_1$ and the lattice constant $a$ are chosen to be unity without any loss of generality. This sets the scales for energy and length respectively.

We have also, in some cases, studied the effect of  including simple three body bond-angle dependent potentials in order to introduce an energy cost for bond bending. These interactions are very well documented in the literature mostly on the system like graphene~\cite{graphene1,graphene2}. To model bond bending we take the Kirkwood~\cite{kkbend2} model, $$\mathcal{H}_{bend}=\frac{k_b}{2}\sum_{\langle i\alpha,j\gamma,k\delta\rangle}\left(\Delta\theta_{i\alpha,j\gamma,k\delta}\right)^2 .$$ which in the small oscillation approximation can be written as 
\begin{eqnarray}
\mathcal{H}_{bend}&\simeq\frac{k_b}{2}\sum_{\langle i\alpha,j\gamma,k\delta\rangle}[\cot\theta_0 (\hat{\mathbf{R}}_{j\gamma,k\delta}\cdot\mathbf{u}_{j\gamma,k\delta}\nonumber\\ 
&+ \hat{\mathbf{R}}_{i\alpha,k\delta}\cdot\mathbf{u}_{i\alpha,k\delta}) -\frac{1}{\sin\theta_0}(\hat{\mathbf{R}}_{i\alpha,k\delta}\cdot\mathbf{u}_{j\gamma,k\delta}\nonumber\\
&+\hat{\mathbf{R}}_{j\gamma,k\delta}\cdot\mathbf{u}_{i\alpha,k\delta} ) ]^2
\label{kkwood}
\end{eqnarray}
where $\theta_0$ is the equilibrium angle and angular brackets denote  triples of particles where $i\alpha$ and $j\gamma$ are both nearest neighbours of $k\delta$.

\begin{figure}[ht]
\includegraphics[scale=0.25,trim=1cm 0 0 0]{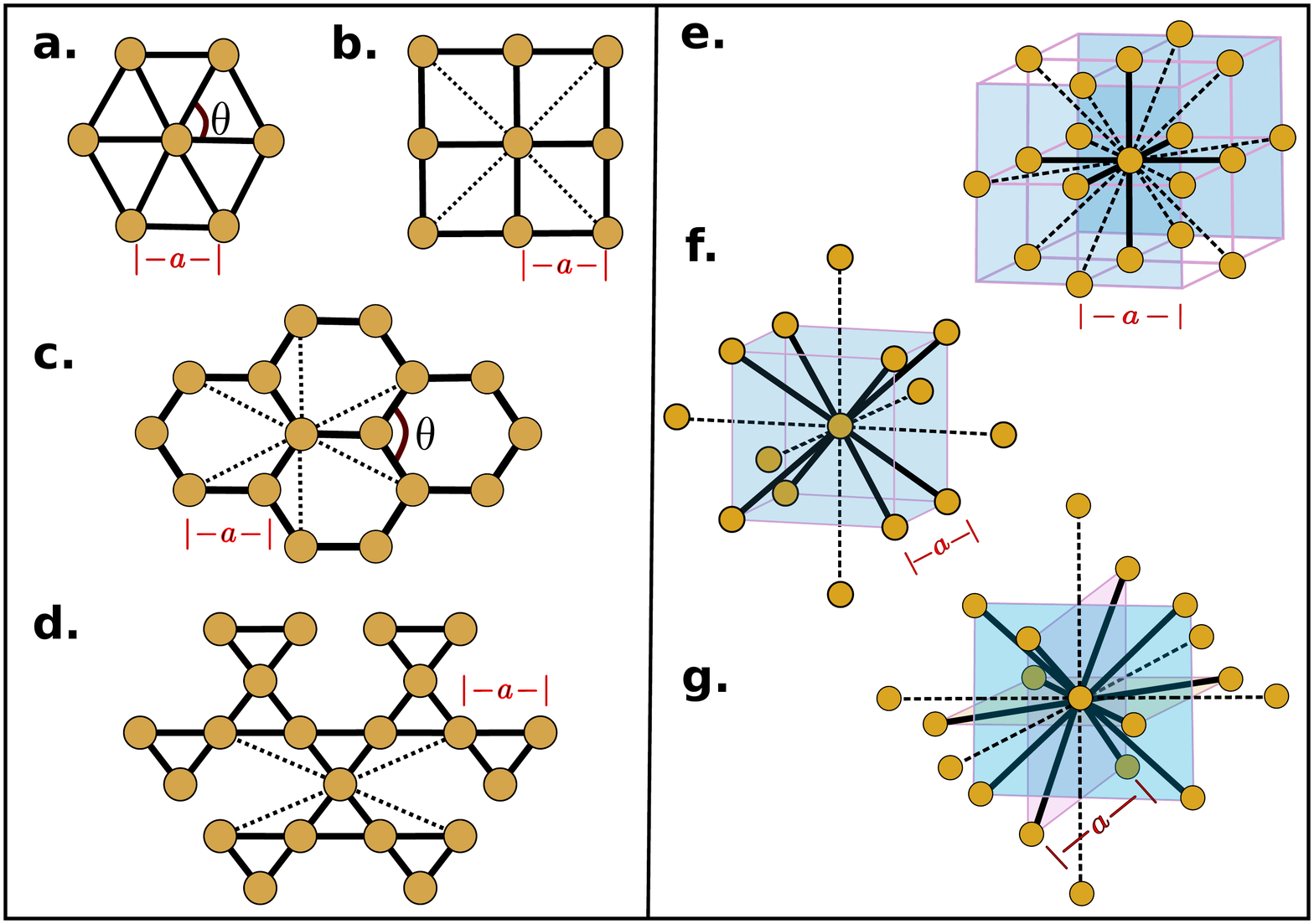}
\caption{Schematic diagram showing the triangular (a), square (b), planar honeycomb (c), kagome (d), simple cubic (e), body centered cubic (f) and face centered cubic (g) lattices. The nearest neighbor bonds are shown in bold while the next nearest neighbor bonds, whenever present, are drawn using  dashed lines. The parameter $a$ is the lattice constant, chosen to be unity. The equilibrium bond angle $\theta_0$ has also been marked for the triangular and honeycomb lattices. In the 3d cases shaded color regions have been added to make the cubic geometry clearer.}
\label{fig0}
\end{figure}

In the following section we discuss our results for specific lattices in 2d and 3d. In Fig.~\ref{fig0} we have shown these lattices schematically and indicated the bonding interactions that we have considered.


\section{\label{results}Results}
We are now in a position to use the methods described in Sec.~\ref{theory} to obtain the statistics of coarse grained non-affine and affine displacements  of  particles interacting through the Hamiltonians presented in Sec.~\ref{models}  ~for a collection of 2d and 3d lattices (see Fig.~\ref{fig0}). As discussed above, the statistics of $\chi$ can be obtained once one has knowledge of the matrix $\mathsf{PCP}$. The projection matrix $\mathsf{P}$ only depends upon the reference position of particles in the lattice and can be constructed easily. The covariance matrix $\mathsf{C}$ can be calculated using Eq.~(\ref{eq:cmat}) once one knows the dynamical matrix $D(\mathbf{q})$. For the harmonic interactions with nearest (and next nearest) neighbours we can compute $D(\mathbf{q})$ in a straightforward manner for all lattices; the results are listed in Appendix~\ref{dynmat}.  The probability distribution for $\chi$ can then be obtained using the eigenvalues of $\mathsf{PCP}$. We have also checked our results by directly simulating the model systems using standard molecular dynamics in the canonical ensemble~\cite{frenkel} as implemented in the LAMMPS simulation package~\cite{LAMMPS}. 
All our results scale linearly with temperature $k_B T = \beta^{-1}$, where $k_B$ is the Boltzmann constant. We have used different temperatures for different lattices only for ease of presentation. 

\begin{figure}[ht]
\includegraphics[scale=0.3]{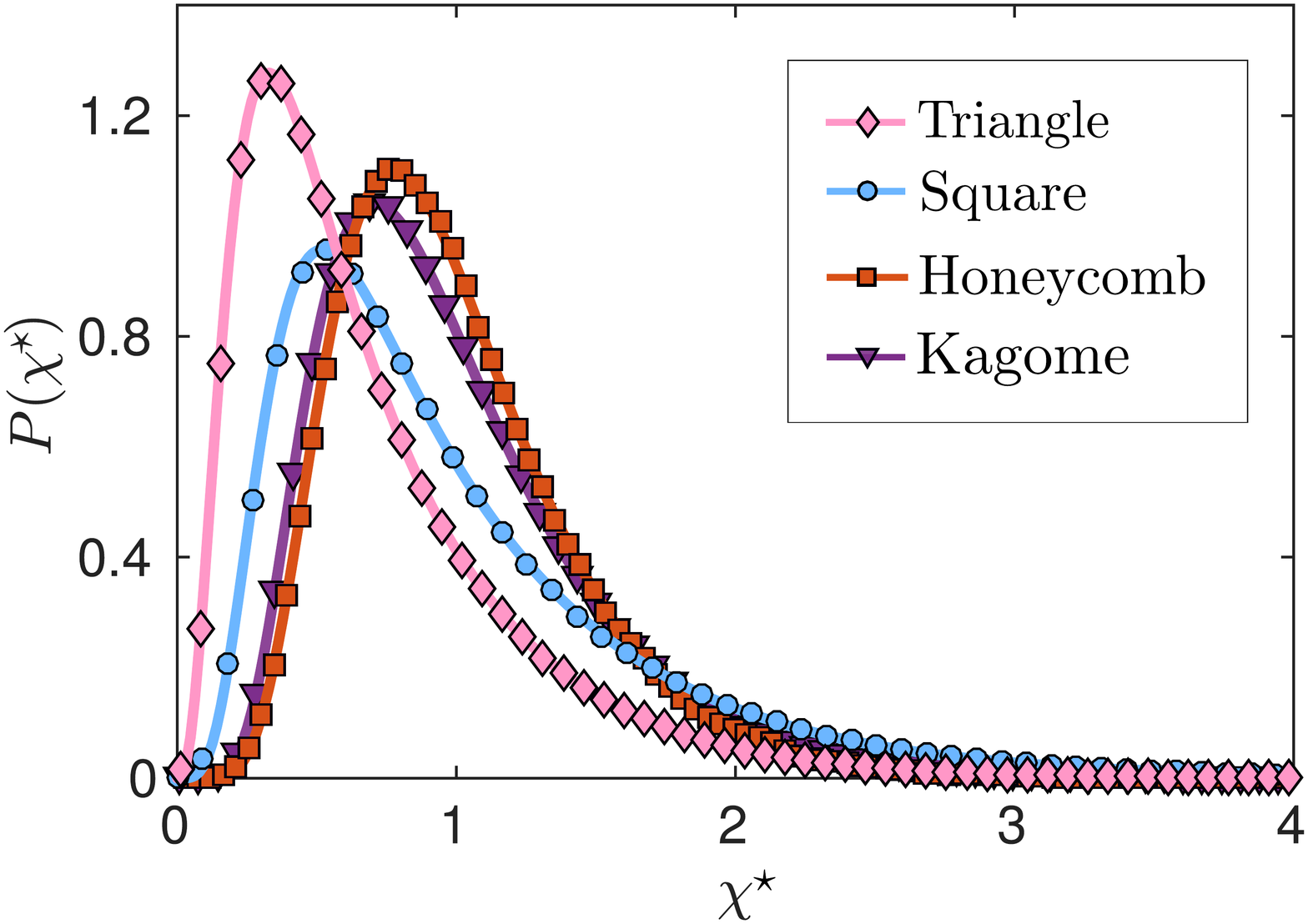}
\caption{ Scaled distibution $P(\chi^\star)$ for all 2d lattices where $\chi^\star = \chi / \langle \chi\rangle$. The solid colored lines are from our analytic calculations and the points are simulation results (using $N=1024$ except for Honeycomb, where $N=512$ and Kagome, where $N=300$). Triangular: light pink with $k_b=0$; Square: sky blue, $k_2 =0.5$; Honeycomb: brown, $k_2=0.5$ and Kagome: purple, $k_2 =0.5$. The distribution for square and triangle plotted here is for the smallest coarse-graining volume used in Sections~\ref{tri} and \ref{sqa}. Whereas, for other lattices coarse-graining volume is same as shown in Fig~\ref{fig5} and \ref{fig7}.
}
\label{fig1}
\end{figure}
\begin{figure}[ht]
\includegraphics[scale=0.3,trim=1cm 0 0 0]{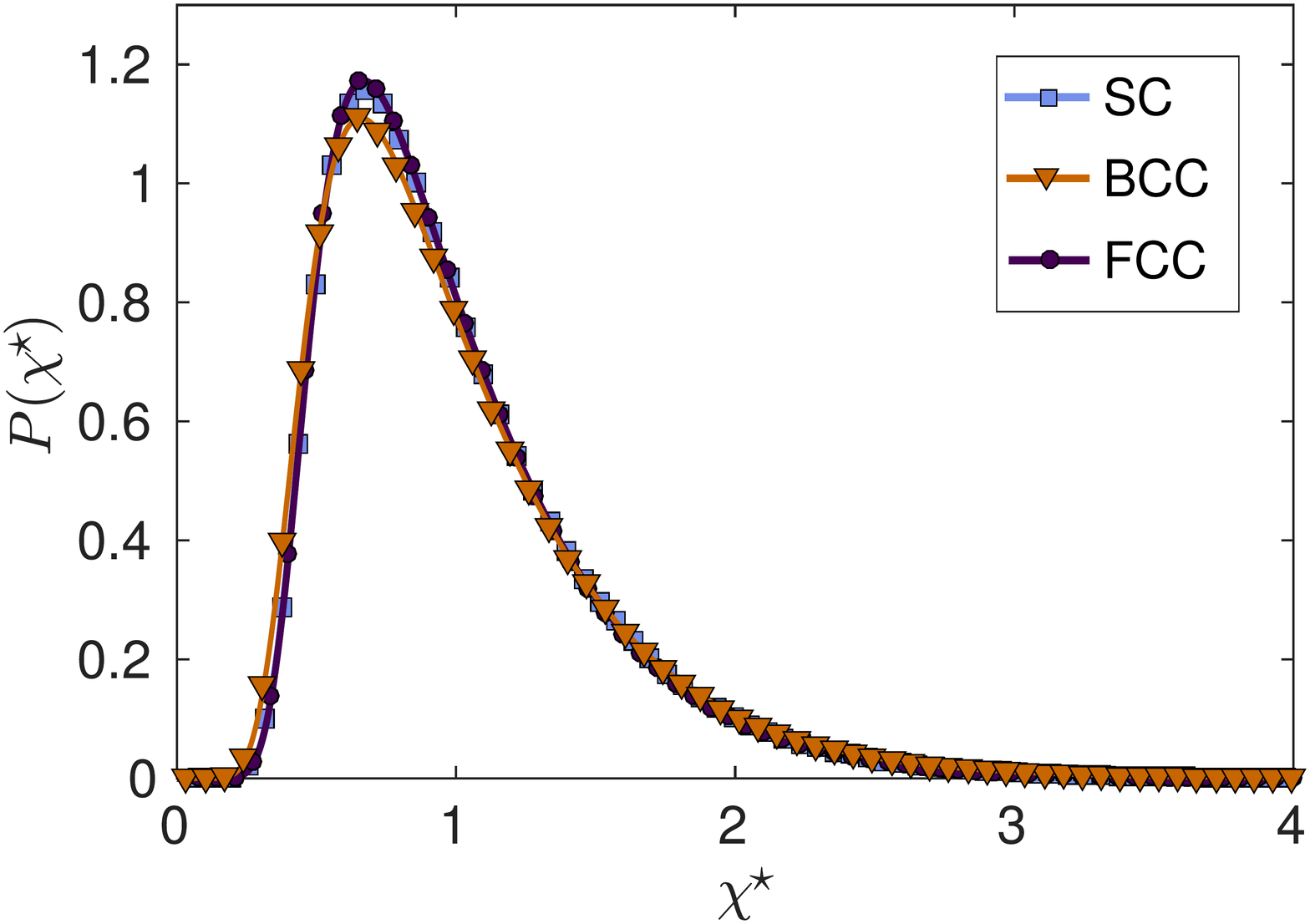}
\caption{Scaled distibution $P(\chi^\star)$ for all 3d lattices, where $\chi^\star = \chi / \langle \chi\rangle$. The meaning of the symbols is the same as in Fig.~\ref{fig1}. The distribution is plotted for SC: sky blue, $k_2=0.5$, $N=1000$.  BCC : brown, $k_2=0.5$, $N=2000$. FCC: purple, $ k_2= 0.75$, $N=4000$. The coarse-graining volume is described in the text.
}
\label{fig2}
\end{figure}

Our results for $P(\chi)$ obtained by numerically inverting $\mathit{\Phi}_{\chi}(k)$ are shown in Fig.~\ref{fig1} and Fig.~\ref{fig2}  together with the results from direct simulations.  All the averaging is done over at least $1000$ well equilibrated and uncorrelated configurations. Needless to say, the agreement is perfect as expected.

Once our formalism is thus established for all the systems considered in 2d and 3d, we turn to each lattice in detail below. We show that using our method  one can find the most prominent non-affine displacement modes (eigenvectors of $\mathsf{PCP}$) for any lattice. Often these modes turn out to be precursors for the most commonly observed defect structures for a given lattice system. 
We note that the relative probabilities of different non-affine modes also depend on the lattice and the interactions and can be easily captured using our approach. 
\begin{figure}[t]
 \includegraphics[scale=0.3]{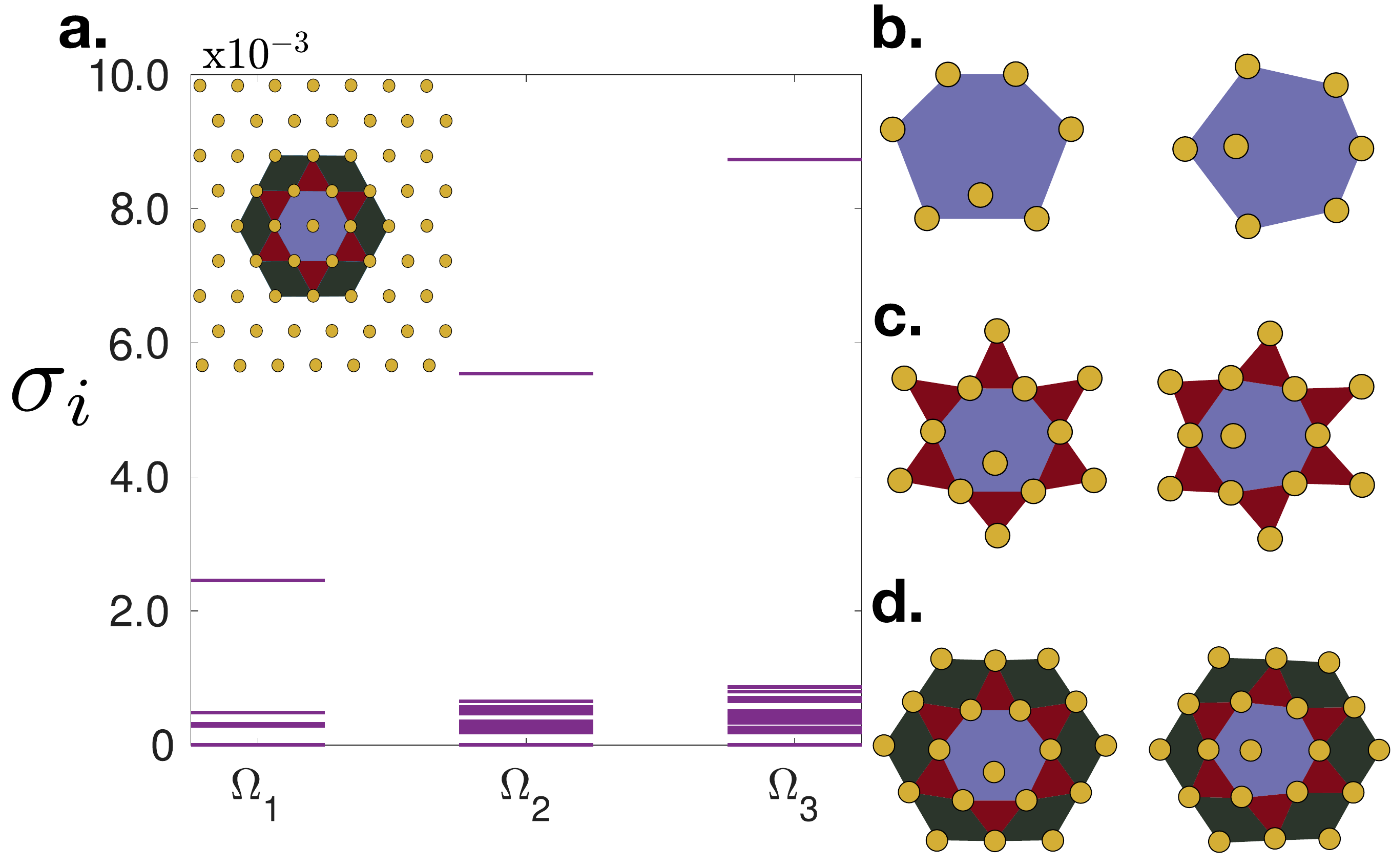}
\caption{Non-affine modes for triangular lattice with different sizes of the coarse graining volume $\Omega$ at inverse temperature  $\beta = 1000$  with no next nearest neighbor bonds. (a) The spectrum of the eigenvalues of $\mathsf{PCP}$ is shown for three different choices of $\Omega$ (inset), which consists of all particles within the first ($\Omega_1$ - light blue), second ($\Omega_2$ - magenta) and third ($\Omega_3$ - dark green) nearest neighbor shells. The reference positions of particles are shown by small yellow circles. The horizontal lines show the eigenvalues. Note the large gap between the largest eigenvalue and the rest of the spectrum. (b) The two degenerate eigenvectors corresponding to the largest eigenvalue of $\mathsf{PCP}$. Note that a nearest neighbor bond is being stretched and a next near neighbor bond nearly perpendicular to it has been shortened. This mode is same as the one discussed in Ref.~\cite{sas2}. This displacement tends to replace the six-fold neighborhoods by two five- and two seven-fold neighbors producing a tightly bound dislocation--antidislocation pair. (c) and (d) show that increasing $\Omega$ does not affect the nature of this mode.}
\label{fig4}
\end{figure}

\subsection{\label{tri}Triangular}
The triangular lattice is the only close packed structure observed in 2d~\cite{CL}. It has just one basis particle per site. We have previously established that the most prominent non-affine mode, i.e.\ the eigenvector corresponding to the largest eigenvalue of $\mathsf{PCP}$, corresponds to the incipient dissociation of a tightly packed dislocation-anti-dislocation pair~\cite{sas2}. To reach this conclusion we used a coarse-graining volume that included only six nearest neighbor particles (see Fig.~\ref{fig4}). This makes $\mathsf{PCP}$ a $12\times12$ matrix with $4$ zero eigenvalues. The non-zero eigenvalues correspond to the independently fluctuating non-affine modes. We have now extended this calculation to include larger $\Omega$. Our results are shown in Fig.~\ref{fig4}. We recover the two degenerate, non-affine modes with the largest eigenvalue described in~\cite{sas2}. We see that these modes continue to be present if one increases the size of the coarse-graining volume $\Omega_1<\Omega_2<\Omega_3$. At the same time the gap between the first eigenvalue and the others increases significantly. 

It is interesting that as the size of $\Omega$ increases, additional vibrational modes go on to populate the lower eigenvalues, keeping the gap intact. We show later that this phenomenon is quite general and observed for many (but not all) lattices. We will comment further on this observation in the discussion (Section~\ref{discuss}). 

\begin{figure}[t]
\includegraphics[scale=0.3]{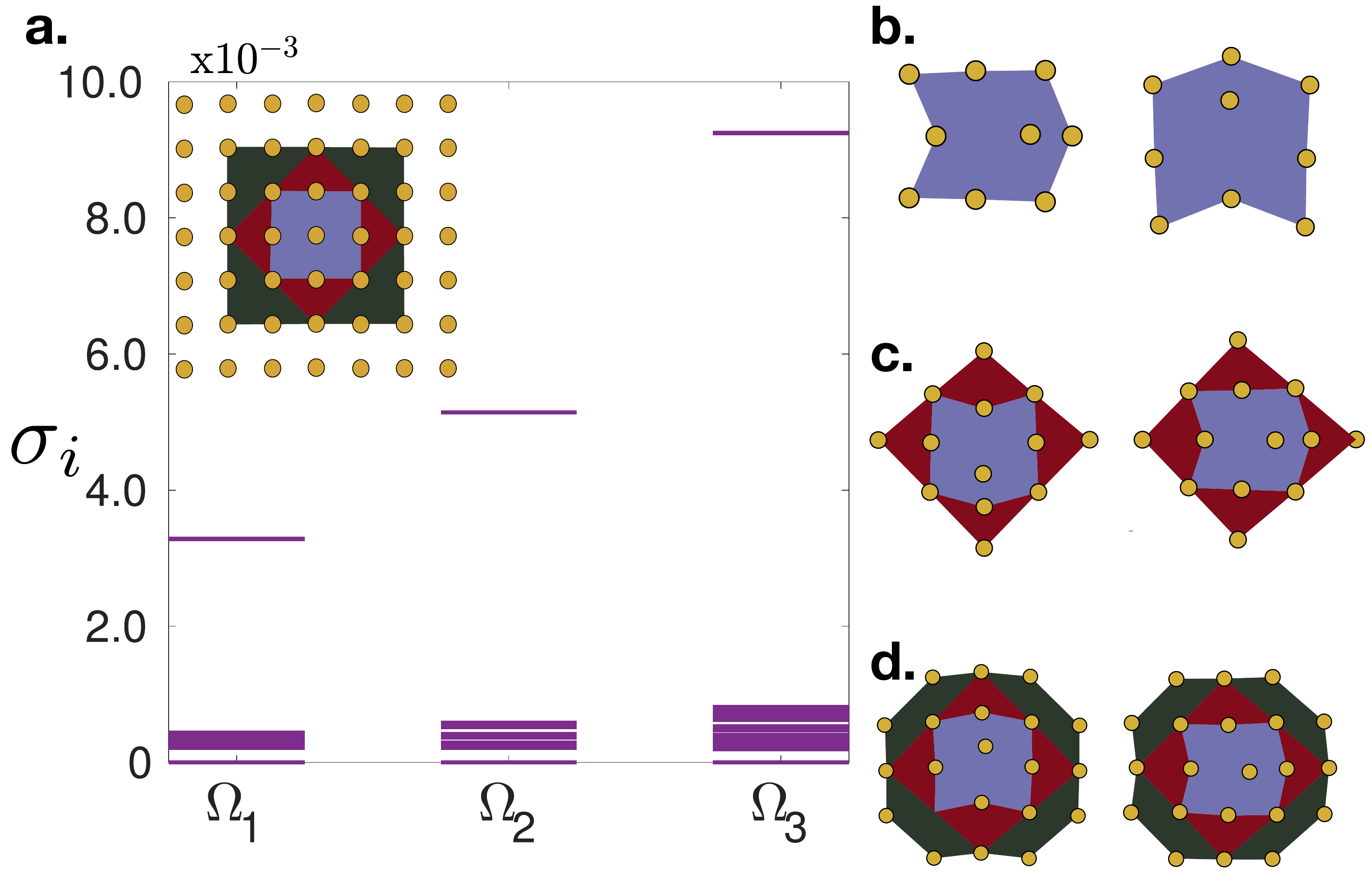}
\caption{Non-affine modes for square lattice with different sizes of the coarse graining volume $\Omega$. The parameters are $k_2 = 0.5$, $\beta = 1000$. (a) The spectrum of the eigenvalues of $\mathsf{PCP}$ is shown for three different choices of $\Omega$ (inset). The color code is the same as in Fig.~\ref{fig4}. (b) The two degenerate eigenvectors corresponding to the largest eigenvalue of $\mathsf{PCP}$. These modes tend to shift a row of atoms relative to adjacent rows. (c) and (d) show that, as in the triangular case, increasing $\Omega$ does not affect the nature of these modes.
%
%
}
\label{fig3}
\end{figure}
\subsection{\label{sqa}Square}
For the square lattice, we need to include both nearest and next nearest neighbor bonds in order to satisfy the Maxwell criteria for stability~\cite{stability}. We have also chosen the smallest $\Omega$ such that all particles to which the central particle is bonded by nonzero interactions are included. This yields an $\Omega$  containing four neighbor and four next neighbor particles. Hence, one obtains $\mathsf{PCP}$ as a $16\times16$ matrix. This has $16$ eigenvalues with eight non-zero values. As the size of $\Omega$ and with it the degree of coarse-graining, is increased one then observes the same effect on the square lattice as in the triangular lattice (see Fig.~\ref{fig3}). Again there is a gap in the eigenvalue spectra between the largest eigenvalue and the rest; this gap increases with the coarse-graining scale chosen. Fig \ref{fig3} shows, in addition to the coarse-graining volumes and the eigenvalue spectra, the softest degenerate eigenmodes. The nature of the mode corresponding to the largest eigenvalue is somewhat different to the triangular lattice case. Instead of introducing defects, it tends to shift the middle row of atoms with respect to its neighboring rows in a direction parallel to the rows. One can easily see that this corresponds to a precursor that can take a square lattice to a triangular one by a shuffle of alternate layers~\cite{poplitrap}. 
\begin{figure}[h]
\includegraphics[scale=0.3]{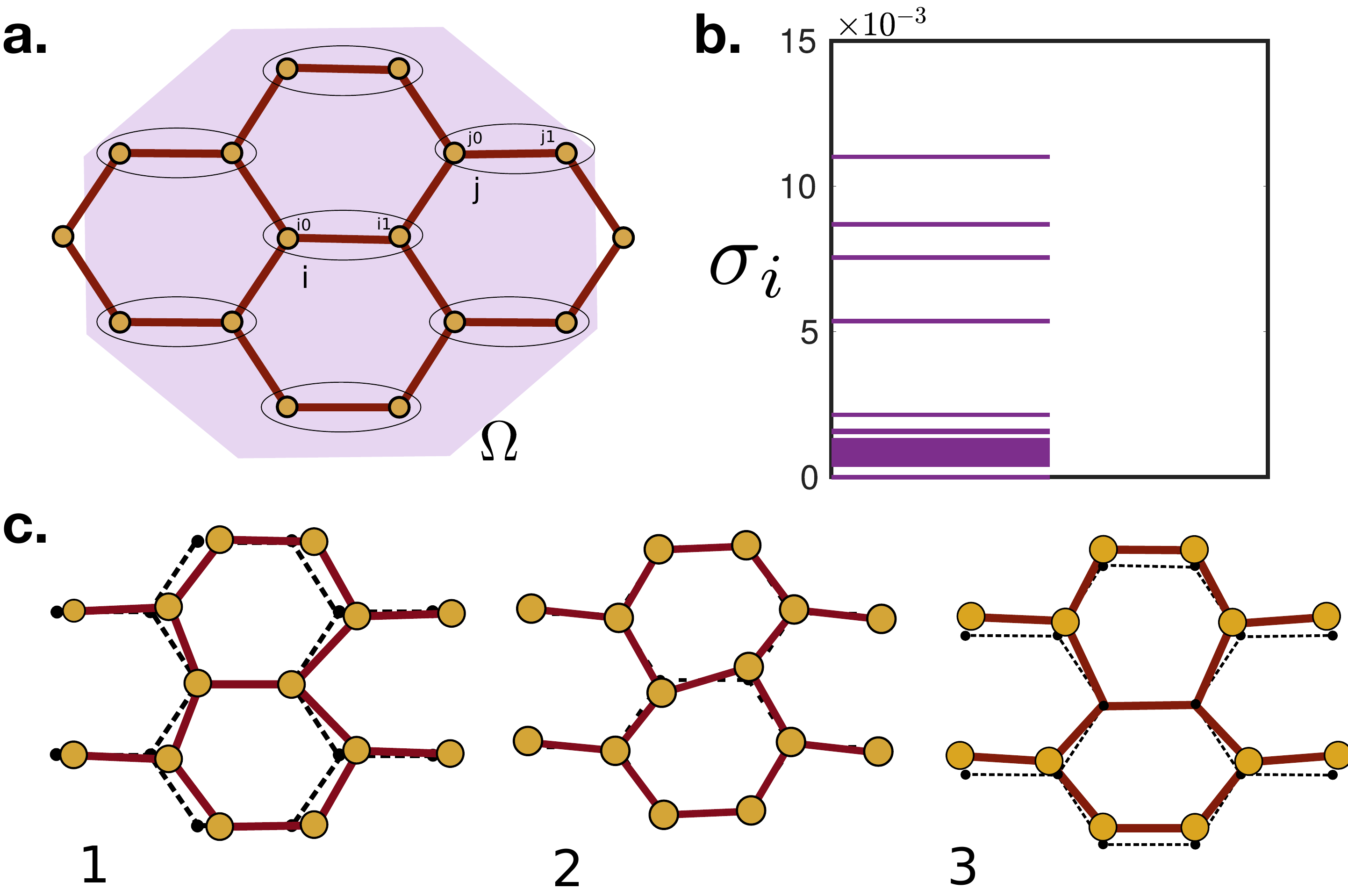}
\caption{Non-affine modes and spectra for the planar honeycomb, with $k_2=0.5$ and $\beta = 500$.
{\bf a.} Schematic of the lattice and the coarse graining volume (pink shaded region) used, with ellipses drawn around each pair of atoms that is in the same basis. {\bf b.} The eigenvalue spectrum. Note that there is no single prominent eigenmode with a large gap as in the triangular and square structures. {\bf c.} Plots of the first three non-degenerate non-affine eigenmodes in the order of prominence (magnitude of eigenvalue). Eigenmode 2 represents an incipient Stone-Wales defect. 
}
\label{fig5}
\end{figure}

\subsection{\label{phc} Planar honeycomb}
The planar honeycomb (or simply honeycomb!) lattice occurs in many condensed matter systems, the most noteworthy being graphene~\cite{graphene1}. A honeycomb lattice is essentially a triangular lattice with a two particle basis. As shown in Fig \ref{fig5}, a cell (ellipse) indexed $i$ has two particles labeled $\alpha \in \{0, 1\}$. Each basis particle has three neighbors and six next nearest neighbors. The coarse-graining volume for a honeycomb lattice Fig.~\ref{fig5}, is constructed as mentioned in Section~\ref{theory} with $\mathfrak{r}$ the next nearest neighbour distance, i.e.\ $ \sqrt{3} a$. Thus, $\Omega$ consists of a total of 17 pairs of particles. (The two particles in the basis each have 3 nearest neighbours and 6 next nearest neighbours, giving 18 particle pairs; excluding from this the double-counted pair of basis atoms yields 17 pairs.) As per the prescription, $\mathsf{P}$ can be constructed and will be a $34\times34$ dimensional matrix. The dynamical matrix is a $4\times4$ matrix with two eigenvalues corresponding to the acoustic branches and the other two to the optical ones. These eigenvalues and eigenvectors of the dynamical matrix are used to calculate the covariance matrix $\mathsf{C}$ as shown in Eq.~(\ref{eq:cmat}). Our experience with the triangular and square lattices shows that increasing the size of $\Omega$ does not influence the nature of the most prominent non-affine modes, although it does considerably increase computational complexity. 

The planar honeycomb structure has been studied in detail in an earlier publication~\cite{sas3}. We include here some of those results for completeness. The probability distribution of $\chi$ is shown in Fig. \ref{fig1} together with the results of other lattices. Fig. \ref{fig5} shows the non-affine eigenvalue spectra. We observe that, in contrast to the triangular and square lattice, there is no clear gap between the largest eigenvalue and the others. We suspect that the presence of optical modes produces an eigenvalue spectrum that does not have pronounced gaps between different modes. It has also been shown in~\cite{sas3} that the nature of this spectrum remains unaffected if one softens the lattices by reducing the value of spring constant $k_2$. In fact, as one softens the lattice these eigenvalues grow without bound, producing more non-affinity in the system. This is obvious because $\mathsf{C}$ is proportional to the lattice Green's function which itself diverges when the spring constant vanishes.

Eigenvectors of $\mathsf{PCP}$ corresponding to the first three largest eigenvalues are plotted in Fig.~\ref{fig5}. Intriguingly, the second mode represents the precursor to a Stone-Wales (SW) defect~\cite{SW}. In SW a central bond flips by $90^\circ$ creating pentagonal and heptagonal voids. One of the questions that was not addressed in~\cite{sas3} was what, if any, is the effect of introducing a bending rigidity~\cite{kkbend1, kkbend2} to the bonds? We take up this issue below. 
\begin{figure}[h]
\includegraphics[scale=0.3,trim=0.5cm 0 0 0]{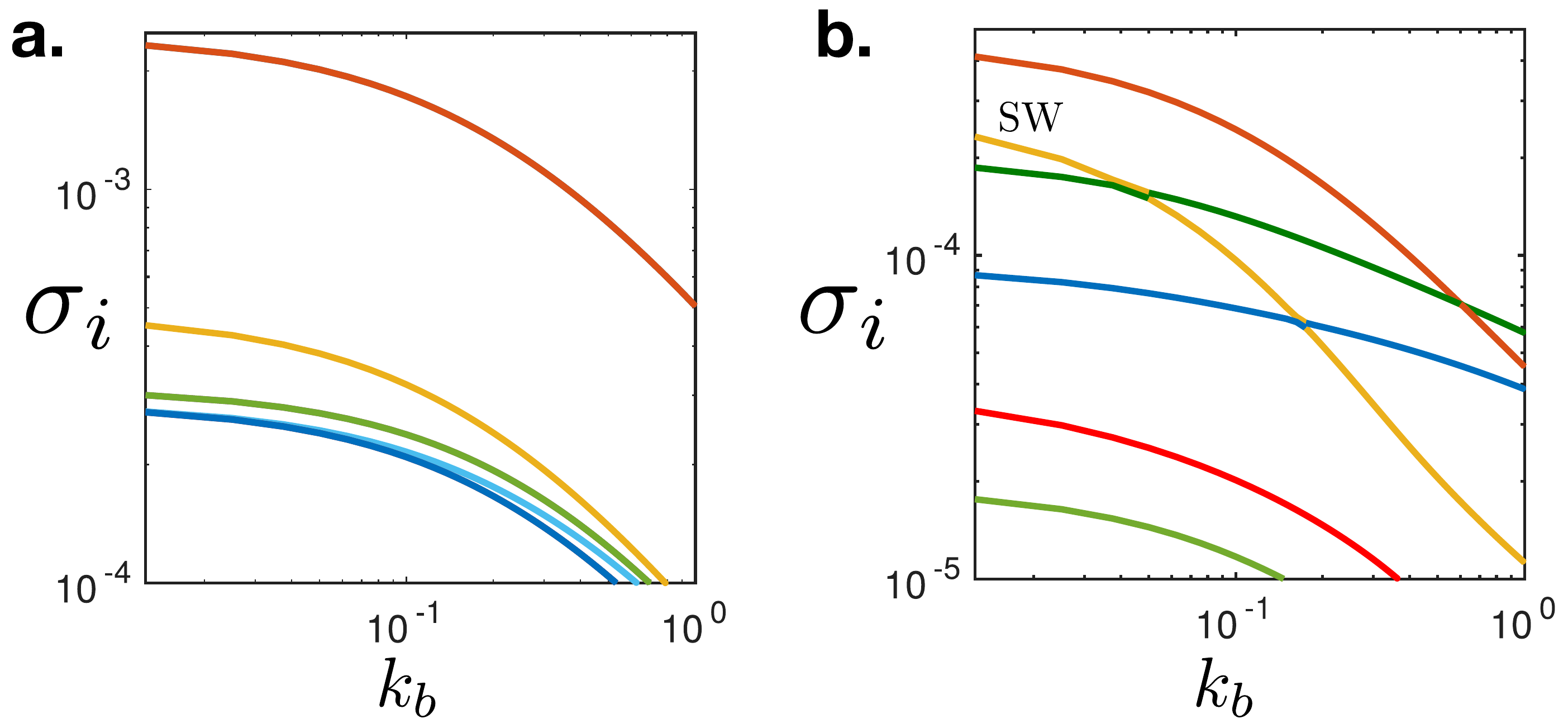}
\caption{Plot of the eigenvalues $\sigma_i$ of $\mathsf{PCP}$ (colored lines, top panel) as a function of the bond angle rigidity parameter $k_b$, showing the  effect of including bending rigidity of bonds in the triangular {\bf a.} and planar honeycomb {\bf b.} structures. While the relative prominence of the modes is unaffected in the triangular lattice except for the breaking of degeneracy of some low probability modes, in the honeycomb lattice the mode corresponding to the SW defect precursor (yellow) is strongly suppressed with increasing $k_b$ (see text).
}
\label{fig6}
\end{figure}

Bond bending rigidity may be modeled as a three body potential given by Eq.(\ref{kkwood}).
The dynamical matrix corresponding to bond bending can be obtained in closed form (see Appendix~\ref{dynmat}). The total dynamical matrix, which is the sum of the dynamical matrices for bond stretching and bending, is  used to calculate $\mathsf{C}$.

Fig.~\ref{fig6} shows how the spectrum of non-affine (non-zero) eigenvalues of $\mathsf{PCP}$ changes upon increasing the bending constant $k_b$. We have included a similar calculation for the triangular lattice for comparison. Because the triangular lattice is isotropic, addition of a bond bending cost only stiffens the lattice and decreases the eigenvalues and consequently their sum, $\chi$. We see that 5-7 defect precursor modes as discussed above continue to be the most prominent modes in the system. Less importantly, the addition of bond bending also breaks the degeneracy of some of the modes corresponding to small eigenvalues.

In contrast to the triangular lattice, the relative prominence of non-affine modes in the planar honeycomb is strongly dependent on the value of the bending  constant. Fig.~\ref{fig6} shows the first six eigenvalues against $k_b$. Several crossovers among the different modes are visible in these spectra. We notice that the SW mode, which earlier was the second most prominent mode in the system, becomes strongly suppressed as one increases $k_b$. This was to be expected because the SW defect requires that nearest neighbor bonds become flexible. Our projection formalism is hence very general and can pick out the dominant defect precursor modes for arbitary lattice symmetry and interactions.

\begin{figure}[ht]
\includegraphics[scale=0.3]{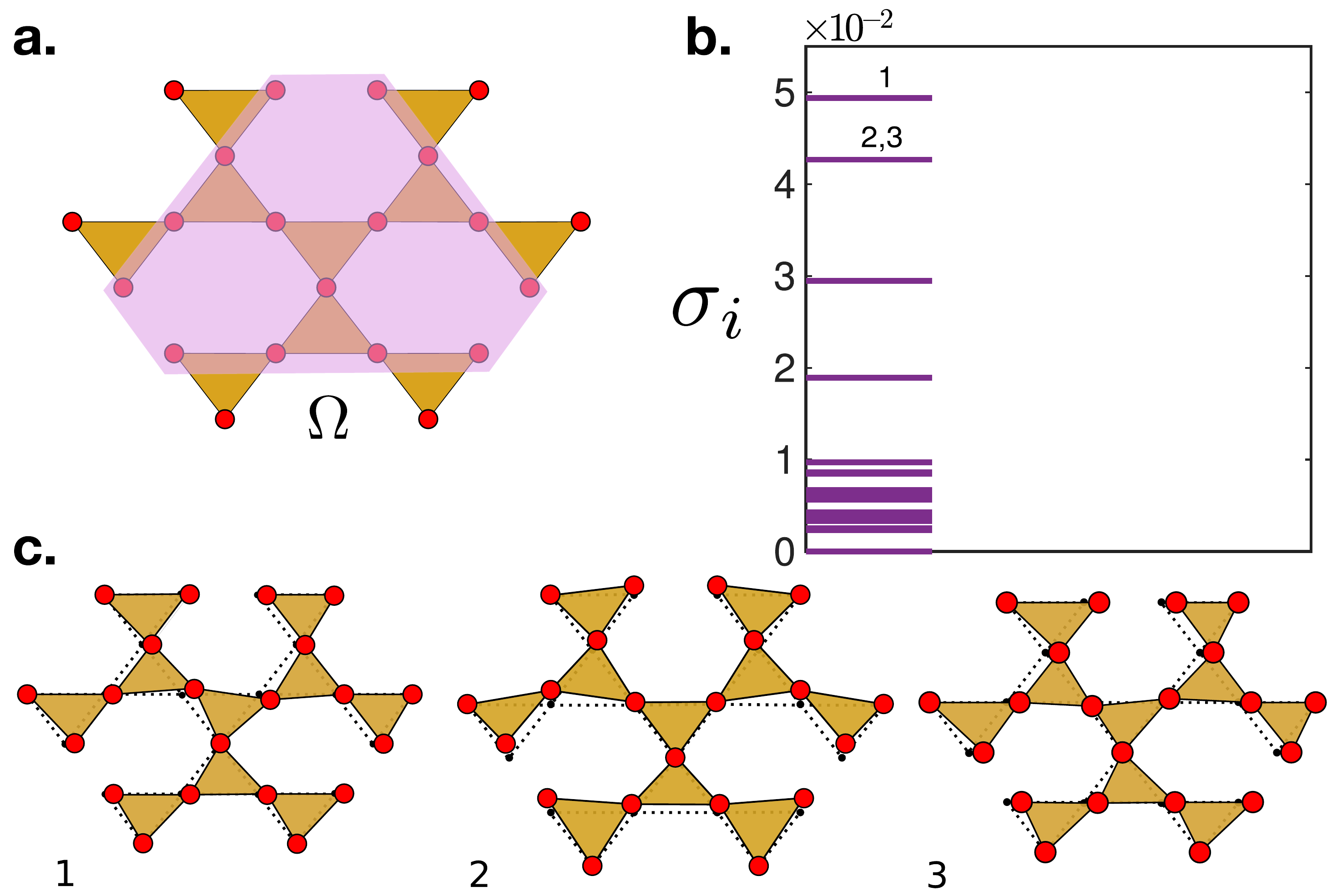}
\caption{{\bf a.} Schematic of the Kagome lattice and coarse-graining volume $\Omega$ (pink shaded region), {\bf b.} the spectrum of non-affine modes and {\bf c.} 1-3, the three most prominent non-affine modes. Parameters used: $ k_2=0.5$, $ \beta = 100$.}
\label{fig7}
\end{figure}

\subsection{\label{kag}Kagome}

The Kagome lattice structure is found in many natural minerals and has interested physicists and chemists because of its unusual magnetic properties~\cite{CL, kagome}. 
Similar to the planar honeycomb, a kagome lattice has a triangular symmetry, but with three basis particles in each cell. Each particle in the cell has four nearest neighbour and four next nearest neighbours. 
The dynamical matrix (Appendix~\ref{dynmat}) $D(\mathbf{q})$ becomes a $6\times6$ matrix and has two acoustic branches and four optical ones. Fig.~\ref{fig7} shows the coarse-graining volume $\Omega$, which contains $21$ pairs of particles up to the next nearest neighbour distance so that $\mathsf{P}$ becomes a $42\times42$ matrix.
Accordingly, $\mathsf{PCP}$ has $38$ non-zero eigenvalues corresponding to non-affine eigenmodes. The probability distribution $P(\chi)$ is shown in Fig \ref{fig1}. Fig \ref{fig7} shows the eigenvalue spectra. Similarly to the planar honeycomb we notice the absence of any large gap among the eigenvalues. The non-affine modes for the largest eigenvalues are shown in Fig.~\ref{fig7}. These modes turn out to be the well known floppy modes~\cite{floppy}.  If the next nearest neighbor bonds are stiffened or bond angle dependent potentials are introduced, the amplitudes of these floppy modes decrease, exactly as in the honeycomb lattice. 

\begin{figure}[t]
\includegraphics[scale=0.3]{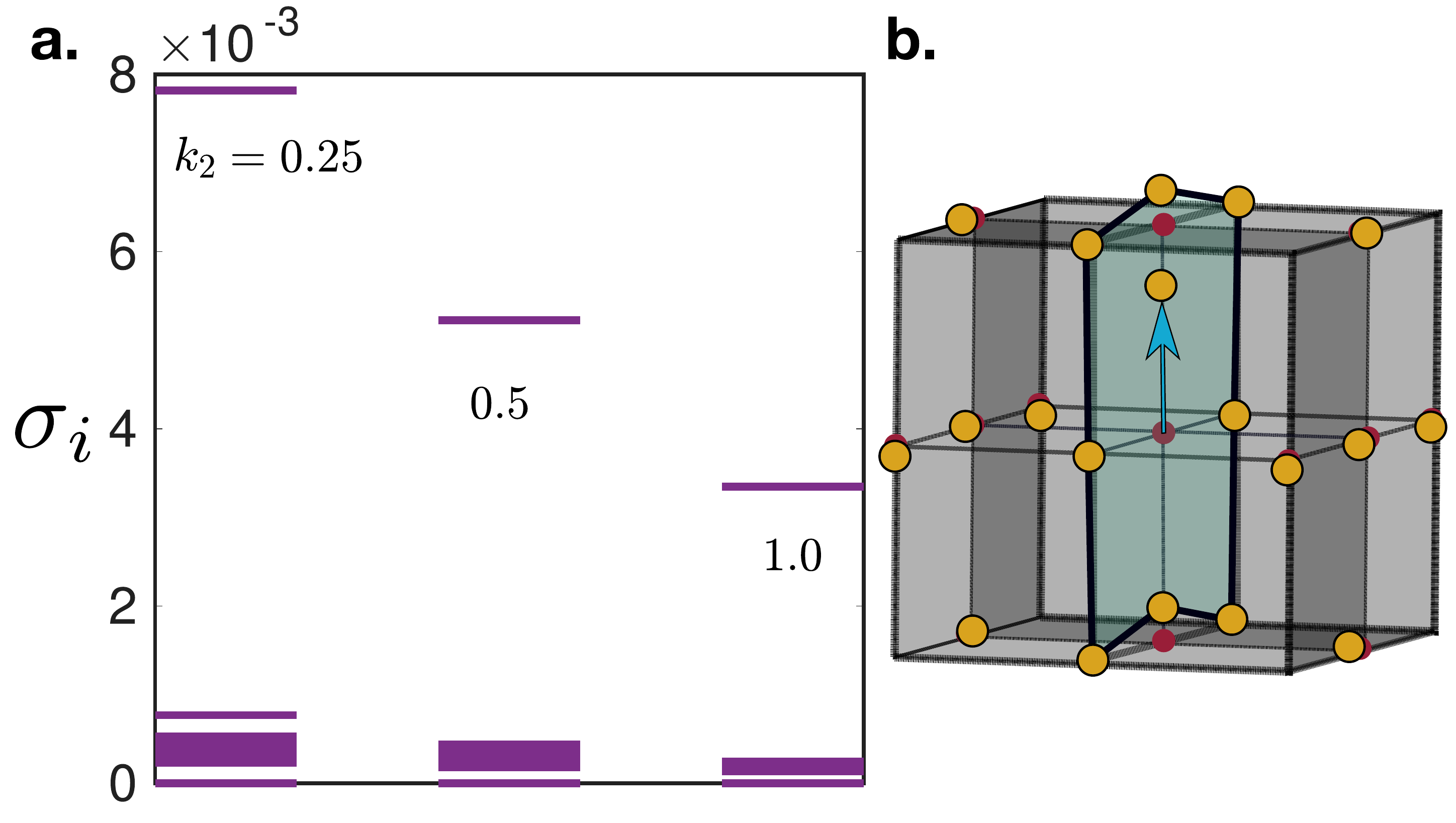}
\caption{{\bf a.} Plot of the eigenvalues of the SC lattice for three different values of $k_2$ : $0.25$, $0.5$ and $1.0$. {\bf b.} The non-affine mode corresponding to the largest eigenvalue for $k_2 = 0.5$ and $\beta = 1000$. Note that this is similar to what is obtained for the square lattice. }
\label{fig8}
\end{figure}

\subsection{\label{sc}Simple cubic}
Our discussion of lattices in three dimensions begins with the simple cubic (SC) lattice, having a single basis atom in a cubic cell with six nearest neighbor and twelve next nearest neighbor particles. We assume that these particles are connected by springs with stiffness constant $k_1$ for nearest neighbors and $k_2$ for next nearest neighbors. The dynamical matrix (Appendix~\ref{dynmat}) can be calculated and has three acoustic branches comprising one longitudinal and two transverse phonon modes. We proceed in a similar fashion as in 2d to calculate $\mathsf{C}$. The projection matrix has $54$ eigenvalues out of which $9$ are zero corresponding to nine affine modes in $3d$. 
Similar to the triangular and square lattices in 2d, we find that a large gap exists between the largest eigenvalue of $\mathsf{PCP}$ and the rest, see Fig.~\ref{fig8}. For the SC, we find that three degenerate modes correspond to this largest eigenvalue, one of which is shown In Fig.~\ref{fig8}. Note that the displacement pattern in the blue shaded plane in SC is similar to that in the square lattice. Indeed the SC lattice may be regarded as a stacking of 2d square lattices. The other two degenerate modes show the same movement in the other two orthogonal planes of the SC. This leads to the interpretation that the most prominent non-affine mode of the SC lattice simply tend to convert the stacked planes from square to triangular symmetry, hence generating 3d close packed structures~\cite{CL}.

The eigenvalue spectra in Fig.~\ref{fig8} also show that $\chi$ decreases as one stiffens the lattice by increasing the stiffness constant $k_2$. However, this increase in stiffness does not affect the qualitative features of the spectrum including the continuing presence of a gap.

\begin{figure}[t]
\includegraphics[scale=0.3]{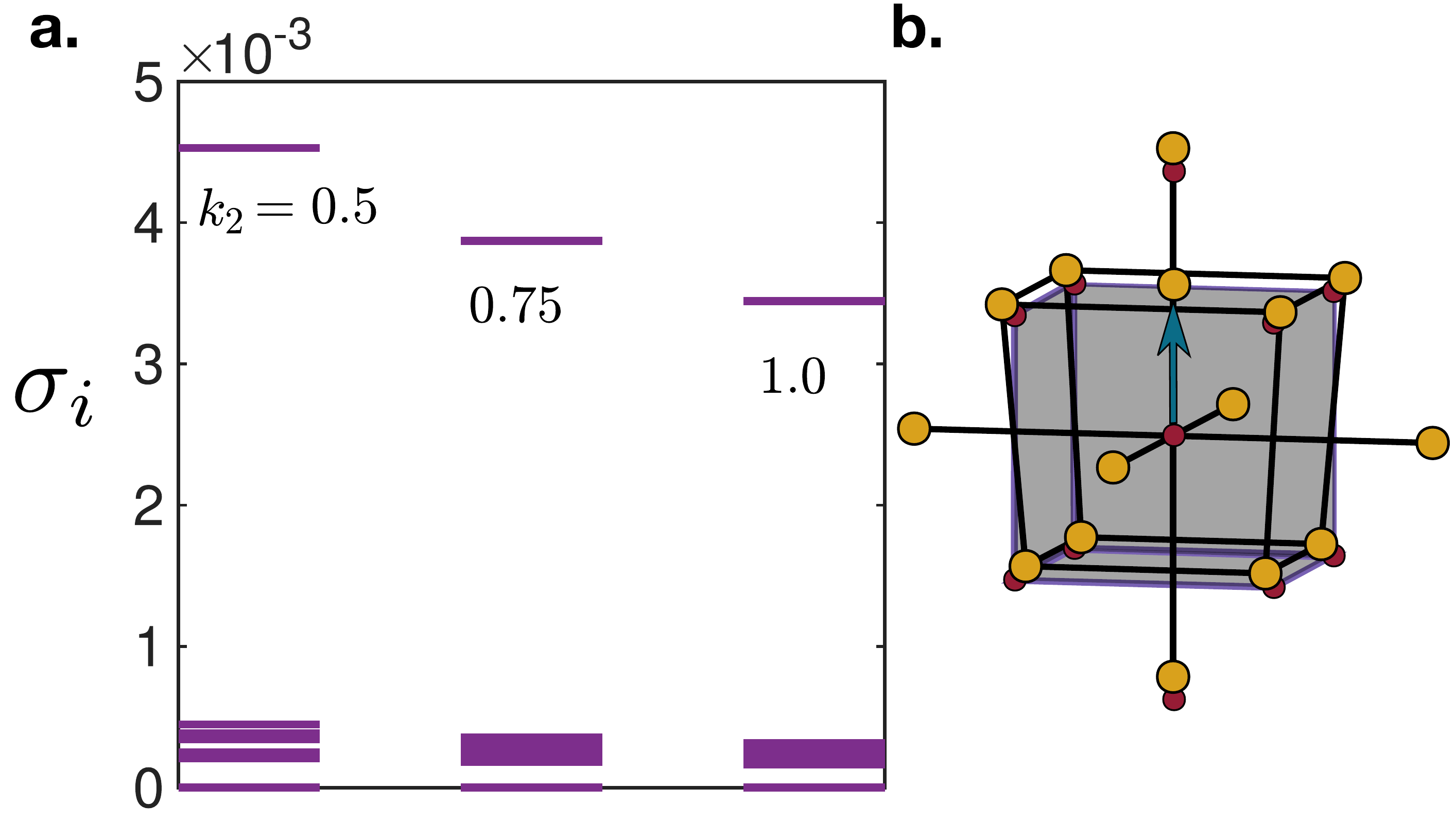}
\caption{{\bf a.} Plot of the non-affine eigenvalues of the BCC lattice for three different values of $k_2$: $0.5$, $0.75$ and $1.0$ at $\beta=1000$. {\bf b.} One of the non-affine modes with the largest eigenvalue for $k_2=0.75$. }
\label{fig9}
\end{figure}

\subsection{\label{bcc}BCC}
The body centered cubic (BCC) lattice may be thought of either as a system with a single-atom basis, or as a SC lattice with a two atom basis~\cite{CL}. For reasons of computational simplicity we choose the former view to construct our coarse-graining volume: this consists of 14 particles with 8 nearest neighbors having bond stiffness $k_1=1$ and 6 next nearest neighbors with bond stiffness $k_2$. 
Since $\Omega$ comprises 14 particles, in three dimensions $\mathsf{P}$  becomes a $42\times42$ matrix, and has  $33$ non-zero eigenvalues corresponding to the non-affine part.

After performing the projection analysis, we find that a gap below the largest eigenvalue is present regardless of the choice of $k_2$. We notice that as for all other lattices discussed above, $\langle\chi\rangle$ decreases with an increase in the stiffness of the lattice, see Fig.~\ref{fig9}. BCC has three degenerate modes related to the largest eigenvalue. One of these is shown in Fig.~\ref{fig9},
where we notice that the centre particle has moved along the $[001]$ direction. The other two modes show a displacement of the centre particle in the two orthogonal directions. These dominant modes
together represent the motion of the body centered particle to one of the six faces of the cubic unit cell, which can be viewed as generating locally a single atomic plane of the FCC lattice by an atomic shuffle. 
%

\begin{figure}[h]
\includegraphics[scale=0.3]{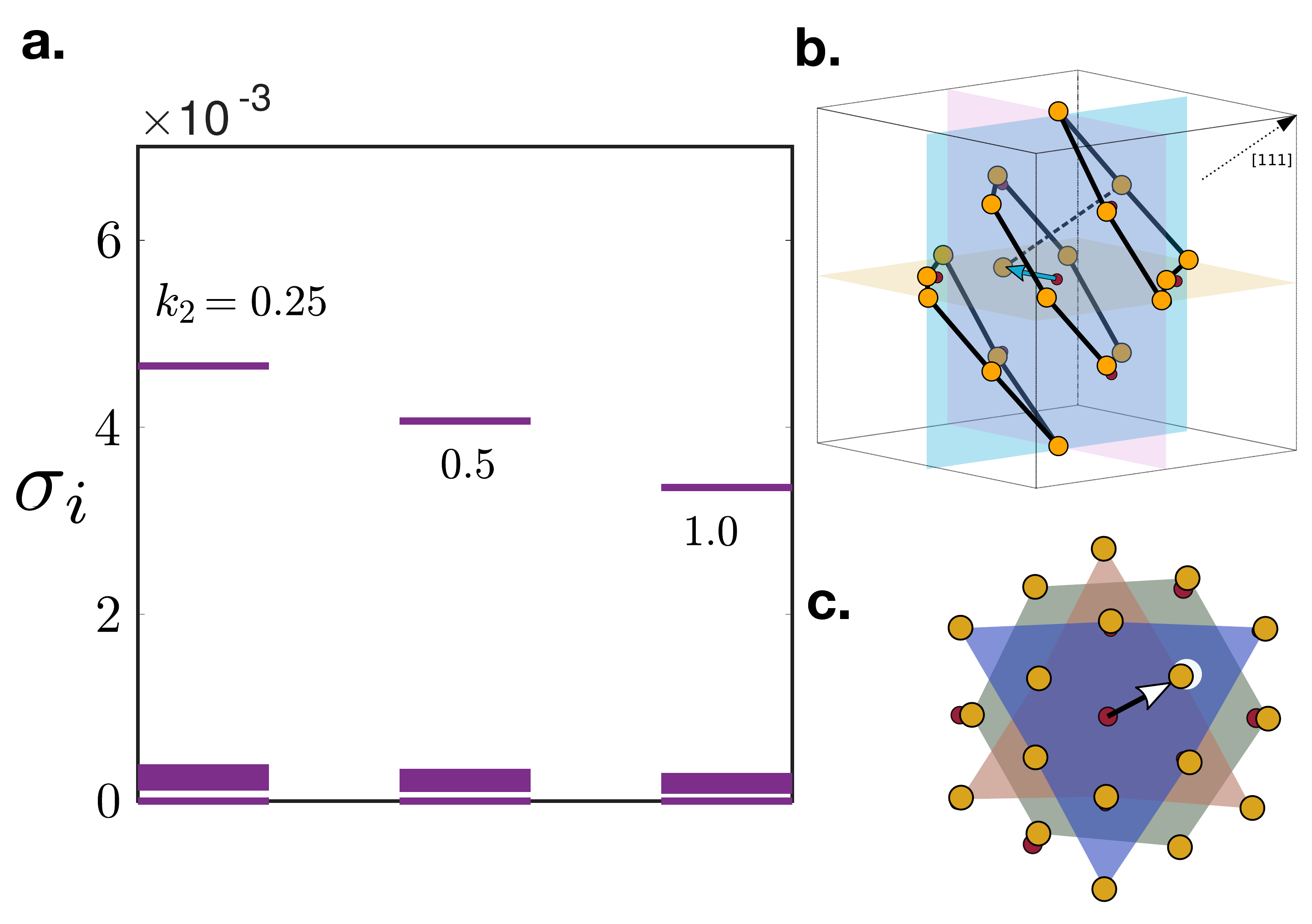}
\caption{{\bf a.}  Plot of eigenvalue spectra of FCC for different choice of $k_2$ : $0.25$, $0.5$ and $1.0$ at $\beta=1000$. {\bf b.} The eigenmode corresponding to the largest eigenvalue for $k_2=0.5$.  {\bf c.} Same as {\bf b.} but viewed from the $[111]$ direction. Notice that the central particle has displaced out of plane and sits below a particle from a different stacking layer, resulting in a stacking fault in the FCC system.}
\label{fig10}
\end{figure}

\subsection{\label{fcc}FCC}
The coarse-graining volume for the face centred cubic (FCC) lattice, we construct around a single atom basis, similar to the BCC case. It consists of 12 nearest neighbor particles and 6 next neighbors.  We thus have $18\times3 = 54$ eigenvalues of $\mathsf{PCP}$, of which 9 eigenvalues representing affine deformations are zero. The eigenvalue spectrum again shows a prominent gap between the three largest degenerate (and mutually orthogonal) eigenmodes and the rest. It is also obvious from the spectra that $\langle\chi\rangle$ decreases as one increases the stiffness of the lattice by increasing $k_2$.

One of the three non-affine modes corresponding to the largest eigenvalue is shown in Fig.~\ref{fig10}. We show later that this mode is a precursor to either a slip or a stacking fault~\cite{CL,rob}. The other two degenerate modes show the analogous deformation in orthogonal directions.

\subsection{\label{coupling}Affine-non-affine coupling}
While the affine and non-affine components of the displacements are orthogonal to each other by construction, they couple at higher order~\cite{sas1}. This has the physical meaning of suggesting that at higher strains, fluctuations which tend to create lattice defects become more probable. While this has been noted in the triangular lattice~\cite{sas1,pnas}, here we undertake a systematic study involving many lattices. 

In Section~\ref{theory} we showed that this coupling is determined by the commutator $[\mathsf{P,C}]$. We have computed this commutator for all the lattices considered in this paper and the results are shown in Table.~\ref{table}. It is interesting to see that open lattices like the planar honeycomb and kagome have larger values of this coupling than the more close packed ones.  
\begin{table}[]
\renewcommand*{\arraystretch}{1.5}
\begin{tabular}{|l|c|c|c|}
\hline
\textbf{Lattice Type}      & $|| C ||$ & $\frac{||[P,C]||}{||C||}$ & Parameters~($k_2$,$k_b$)\\ \hhline{|====|}
2d triangle\hspace{3pt} &    3.874    & 0.031  &0.5, 0 \\ \hline
2d triangle \hspace{3pt}   & 1.290    & 0.013&  0.5, 0.5\\ \hline
2d square \hspace{3pt} & 5.055    & 0.037 &   0.5, 0\\ \hline
2d honeycomb \hspace{3pt}  & 9.362    & 0.160 & 0.5, 0\\ \hline
2d honeycomb \hspace{3pt}  & 4.897    & 0.135 & 0.5, 0.5\\ \hline
2d kagome\hspace{3pt}      & 9.910    & 0.113 &   0.5, 0\\ \hline
3d SC\hspace{3pt}     & 9.361    &0.024 &   0.5, 0 \\ \hline
3d BCC \hspace{3pt}  & 6.994    &0.020 &   0.75, 0\\ \hline
3d FCC\hspace{3pt}   & 7.279   & 0.025 &   0.5, 0\\ \hline
\end{tabular}
\caption{The Frobenius norm (the square root of the sum of the absolute squares of the elements) of the commutator $[\mathsf{P,C}]$, made dimensionless by dividing it by the corresponding norm of $\mathsf{C}$, for a number of lattices in 2d and 3d at $\beta=1$. Corresponding parameter values for stiffness are quoted in the last column. Also note that the norm of $\mathsf{P}$ is essentially the square-root of total number of non-affine modes in each case.
}
\label{table}
\end{table}

\begin{figure}[h]
\includegraphics[scale=0.3]{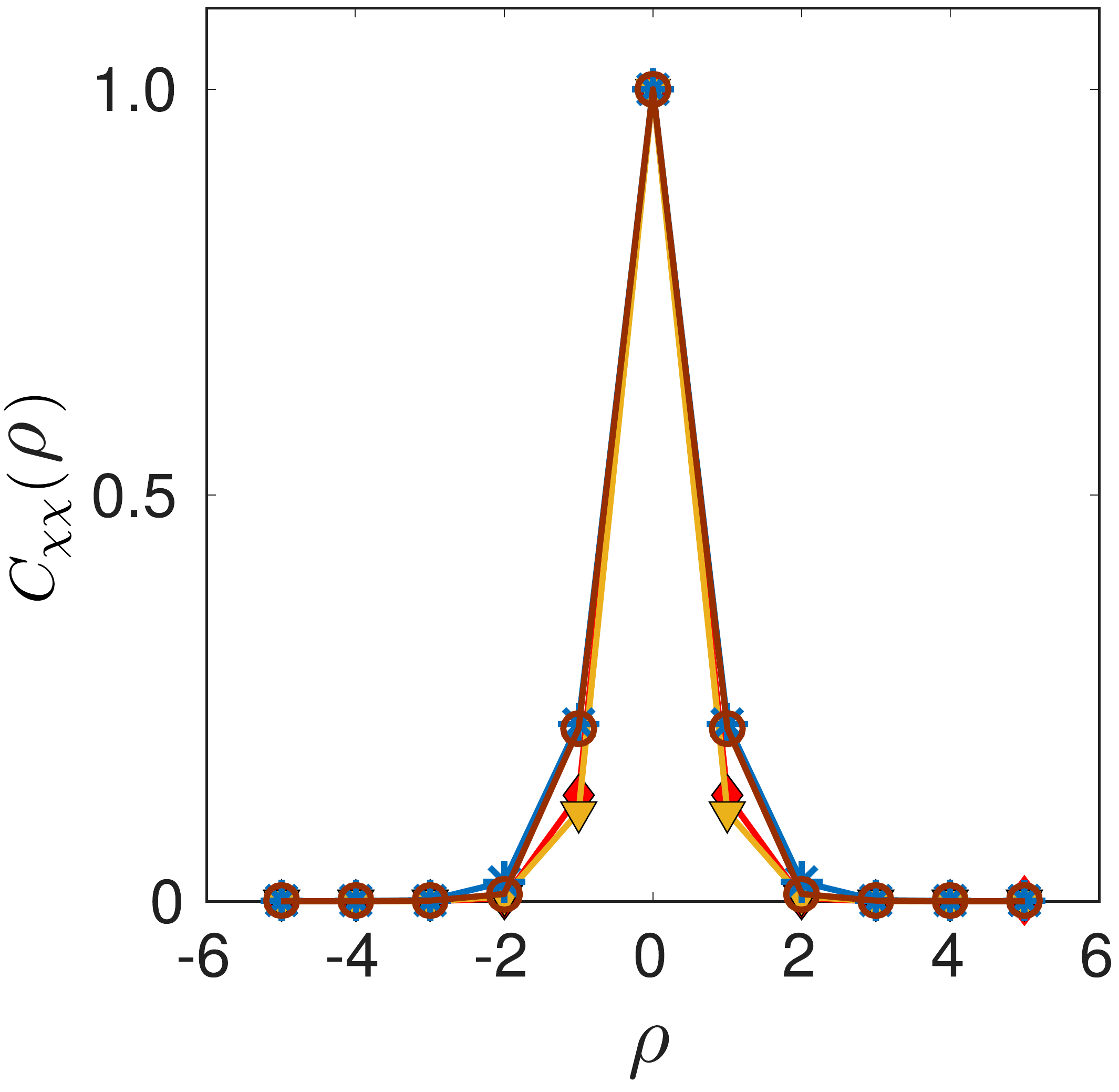}
\caption{Normalized $\chi$ correlations, $C_{\chi\chi}(\rho)=(\langle\chi(0)\chi(\rho)\rangle - \langle\chi(0)\rangle^2)/\left(\langle{\chi(0)}^2 \rangle- {\langle \chi(0) \rangle}^2\right)$, for several different lattices as a function of distance $\rho = {\bf R}\cdot \hat {\bf x}/d_{nn}$ measured in units of the {\em nearest neighbor distance},$d_{nn}$ in the reference lattice along one coordinate axis. Brown and blue are for square and triangle lattices ($d_{nn}=a$); orange and red are for FCC ($d_{nn}=a/\sqrt{2}$) and BCC ($d_{nn}=a\sqrt{3}/2$) respectively.}
\label{fig12}
\end{figure}
\subsection{\label{corr}Spatial correlations}
We now look at two-point spatial correlations of $\chi$ and the affine strains $\mathbf{e}$. These have been extensively studied for two dimensional lattices both numerically~\cite{kerst1,kerst2} and analytically~\cite{sas1,sas2,sas3}. The spatial correlation of the affine strain is important because it offers a way to obtain elastic properties of colloidal crystals from optical microscopy images~\cite{zahn}.  

The spatial correlations of $\chi$ for some of the lattices considered in this paper are shown in Fig.~\ref{fig12} in a single plot. These correlations are nearly isotropic and are plotted as a function of distance $\rho$  measured in the units of nearest neighbour distance $d_{nn}$ along one coordinate axis. The values of $d_{nn}$ for different lattices are mentioned in Fig~\ref{fig12}. We observe that the nature of the correlation function is similar for all lattices. It is a sharply decaying function that essentially vanishes after the second neighbor shell. More quantitatively, we observe that the correlations decay somewhat faster in higher dimensions. 

The spatial correlations for the affine strains may be obtained using the procedure outlined in Section~\ref{theory}. These have a more non-trivial structure. They are anisotropic and can be long-ranged along particular directions~\cite{kerst2}. In the ${\bf q}\to 0$ limit, analytic expressions for these correlation functions can be derived quite easily. For example,  as defined in Eq.~(\ref{eq:straincorr}), the strain correlation functions for the square lattice are, $\beta \langle e_v^2\rangle ({\bf q}) = Q_v/Q$,  $\beta \langle e_u^2\rangle({\bf q}) = Q_u/Q$, and $\beta\langle e_s^2\rangle({\bf q}) = Q_s/Q$ with the abbreviations
\begin{eqnarray}
Q & = & q_{x}^2 q_{y}^2+k_{2} \left(q_{x}^2+q_{y}^2\right)^2+k_{2}^2 \left(q_{x}^2-q_{y}^2\right)^2 \nonumber \\
Q_v & = & 2 q_{x}^2 q_{y}^2+k_{2} \left(q_{x}^2-q_{y}^2\right)^2 \nonumber \\
Q_u & = & 2 q_{x}^2 q_{y}^2+k_{2} \left(q_{x}^4+6 q_{x}^2 q_{y}^2+q_{y}^4\right) \nonumber \\
Q_s & = & \left(q_{x}^4+q_{y}^4\right)+k_{2} \left(q_{x}^2-q_{y}^2\right)^2
\label{2dsqcor}
\end{eqnarray}

%

\begin{figure*}
\begin{center}
\includegraphics[scale=0.3]{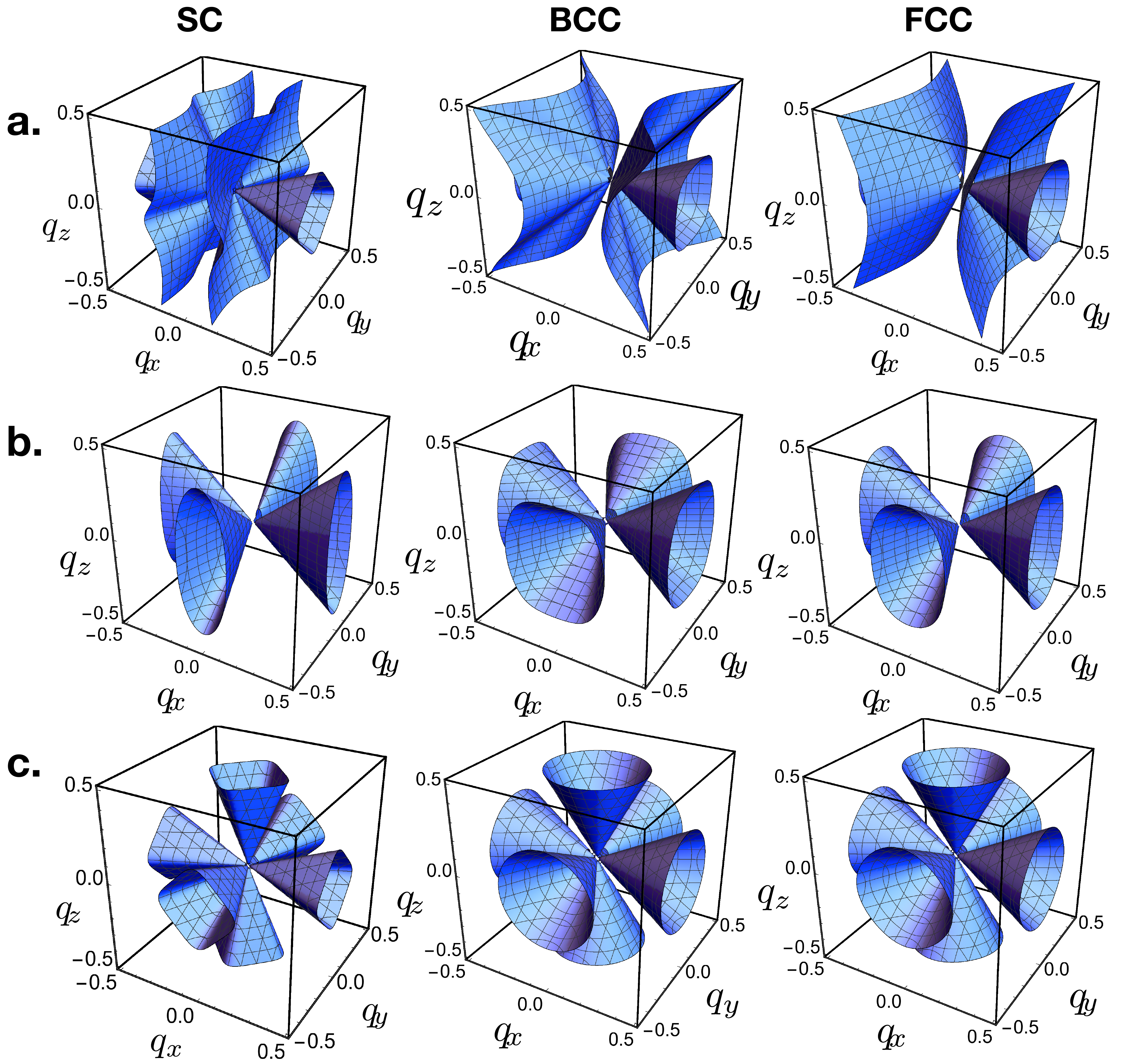}
\end{center} 
\caption{Iso-surfaces for the strain-strain correlation functions in Fourier space and in the ${\bf q}\to 0$ limit for SC, BCC, FCC lattices. Figures shown for {\bf a.} deviatoric ($e_{xx}-e_{yy}-e_{zz}$), {\bf b.} shear ($e_{xy}+e_{yx}$) and {\bf c.} volume ($e_{xx}+e_{yy}+e_{zz}$) strains. The values of the correlations at the iso-surfaces are different in each case and have been chosen for ease of presentation. They are listed in the Appendix~\ref{strain3d} along with the full algebraic expressions used to plot the  iso-surfaces. The other parameters used are $k_2 = 1/6$ and $\beta=1$ throughout.}
\label{fig14}
\end{figure*}

Similar expression for the triangular lattice have already been discussed in~\cite{sas1}. We observe that for $k_2=1/2$, where the square lattice becomes elastically isotropic~\cite{isotropy}, the   
expressions in  Eq.~(\ref{2dsqcor}) differ from those for the triangular lattice only by an unimportant overall factor. The general shape of these correlation functions, viz.\ the ``butterfly pattern", is also similar to results obtained in colloidal glasses using video microscopy techniques~\cite{chikkadi}. 

In three dimensions, the correlation functions are considerably more complicated, although analytic expressions in Fourier space in the small wave-number limit can still be worked out with some effort. The algebraic expressions are given in Appendix~\ref{strain3d} and they are plotted in Fig.~\ref{fig14}. 
%

\section{\label{discuss}Discussion and Conclusions}
\label{discuss}
In this paper we have studied the nature of thermally excited non-affine atomic displacements for a number of crystalline solids in 2d and 3d. We have discovered several features that are common to many lattice systems although close packed and open lattices show somewhat different properties. While in close packed lattices, the contribution to $\chi$ is dominated by a single non-affine mode (or degenerate, symmetry-related class of modes), in open lattices there is no such predominance. Further, in open lattices, the contribution of the different modes is much more sensitive to details of the interactions and they are more strongly coupled to affine fluctuations.  
\begin{figure}[ht]
\includegraphics[scale=0.2,trim=3cm 0cm 0 0]{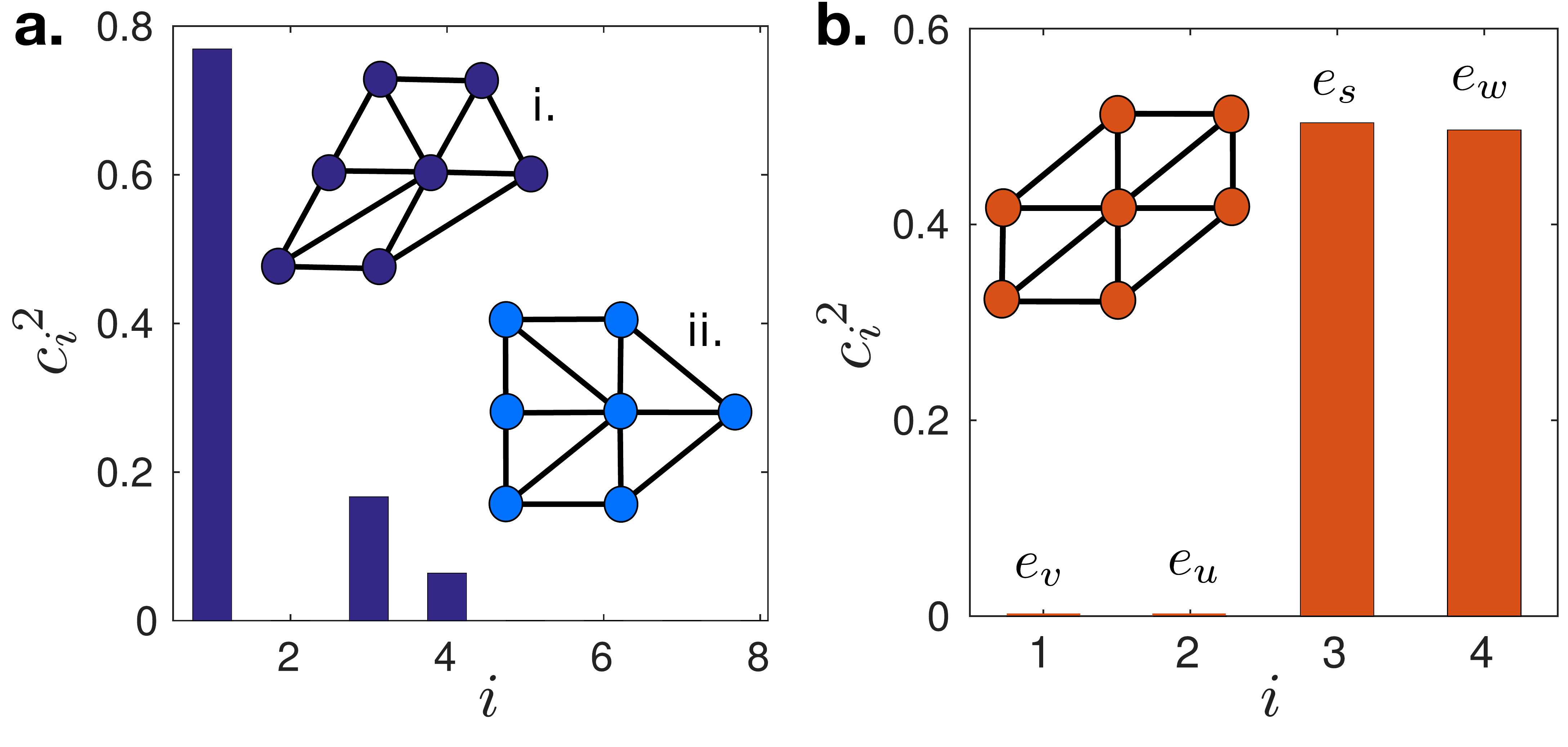}
\caption{{\bf a.} Relative contribution from non-affine modes when a slip is introduced by translating the bottom half of a triangular lattice along a close-packed direction by a lattice spacing.  {\bf b.} The corresponding contribution of the four affine modes. Note that the largest contributions come from shear and rotation. The insets show the configuration of atoms in $\Omega$ after the slip ({\bf a}-i) and the separate non-affine ({\bf a}-ii) and affine ({\bf b}) contributions. See text for details.}
\label{fig16}
\end{figure}

One of the important findings of our earlier work~\cite{sas2,sas3} was that non-affine displacement fluctuations serve as precursors to the formation of defects. In the triangular lattice, in the presence of strain, a dislocation-anti-dislocation pair separates and produces a slip plane~\cite{pnas} that has high values of $\chi$. We end this paper by carrying out a simple exercise in order to better understand the relation between non-affine modes and defects. 

Accordingly, we first consider a triangular lattice where a slip is introduced such that a part of the lattice moves a lattice spacing in a close packed direction, compared to the rest. In the bulk, there is no contribution to $\chi$ as all atoms undergo either no motion or just a uniform translation. Thermal vibrations are neglected in this calculation and those that follow. We choose a coarse graining volume corresponding to the smallest $\Omega$ as shown in Sec.~\ref{results} centered on an atom lying in the slip plane. The $\Omega$ at the interface of the slipped and un-slipped regions is, of course, deformed (see Fig.~\ref{fig16}). This deformation cannot be described by a homogeneous affine transformation of $\Omega$ and therefore contributes to $\chi$. Including thermal contributions would produce a $P(\chi)$ that is identical to the ones calculated in the bulk (Section~\ref{results}), while in the vicinity of the slip, $P(\chi)$ would be displaced to higher $\chi$ values~\cite{pnas}.

One can now project this deformation 
onto the non-affine and affine modes computed from thermal averages of displacements to find the contribution of individual modes to this deformation. In Fig.~\ref{fig16}, we plot the bar-graph of the components ${(c_i)}^2$ obtained by projecting onto the affine and non-affine modes for this deformation, where $c_i$ is the coefficient corresponding to the $i^{th}$ mode in the expansion of the displacement as a superposition of non-affine and affine modes. 
We see that the largest contribution comes from the first two non-affine modes as expected. There is also a non-zero contribution from the affine modes, with the affine and non-affine modes contributing equally overall.
%
%
%
This may be easily understood from the insets shown in Fig.~\ref{fig16} {\bf a} and {\bf b}. In Fig.~\ref{fig16}~{\bf a} (inset i), we show the configuration of particles where the two atoms belonging to the bottom-most row are displaced to the left by a lattice spacing relative to the upper two rows. The total non-affine contribution is shown in inset ii of the same figure. This shows a relative displacements to the right of the middle row consisting of three atoms. On the other hand, the affine contribution to the slip shown in Fig.~\ref{fig16}~{\bf b} (inset) consists of {\em homogeneous} shear and local rotation as verified from the bar-graph. It is clear that the sum of the affine and non-affine displacements gives rise to the slipped configuration shown in Fig.~\ref{fig16}~{\bf a} (inset i). The affine deformation produces an internal shear stress at the slip plane. When a crystal slips in response to an external homogeneous shear, the internal stress cancels the external stress locally. By introducing a finite density of such slip planes, any homogenous stress may be expelled. In Ref.~\cite{pnas} such an expulsion process was shown to lead to yielding of crystalline solids at any shear stress, however small. Since the non-affine strains corresponding to the largest eigenvalues do not depend on the choice of $\Omega$ (see Section~\ref{results}), the mechanism described is quite general.
\begin{figure}
\begin{center}
\includegraphics[scale=0.3,trim=4cm 0 0 0]{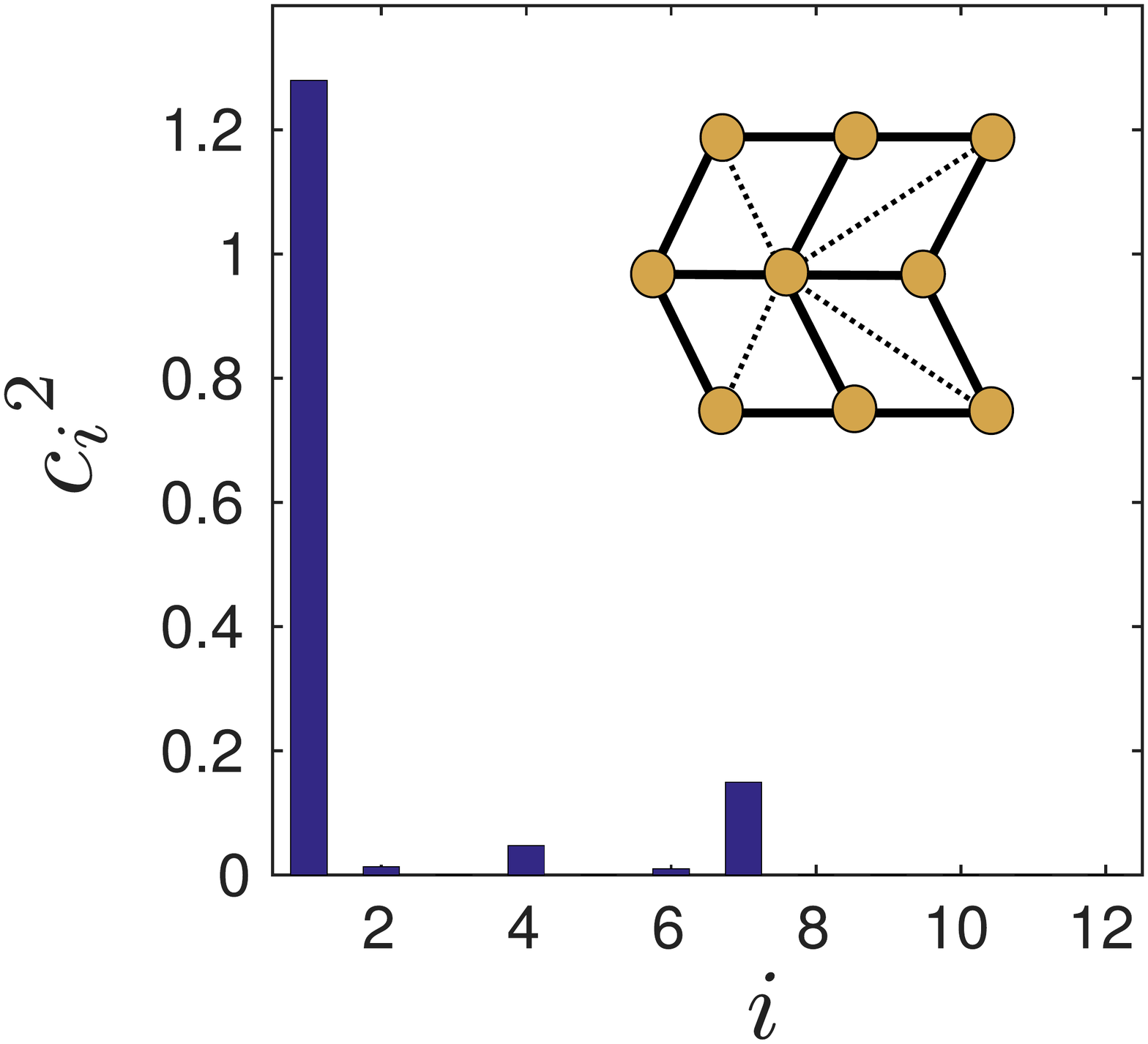} 
\caption{Non-affine contribution in the square lattice when the middle row is displaced by half a lattice spacing. Note that this transformation tends to produce a triangular lattice symmetry starting from the original square lattice.}
\end{center}
\label{fig15}
\end{figure}

We now turn to the square lattice. It is known~\cite{poplitrap} that a square lattice can transform to a triangular lattice by either a homogeneous, affine, shear or by a non-affine deformation where alternate rows of atoms shift by half a lattice spacing together with a homogeneous relaxation of the lattice parameters. We show this deformation in Fig.~\ref{fig15}(inset) and compute the projection onto the non-affine modes. As expected, there is an overwhelmingly large contribution from the non-affine mode with the largest eigenvalue. There is no affine component for this deformation. 

\begin{figure}
\includegraphics[scale=0.25,trim=5cm 0 3cm 0]{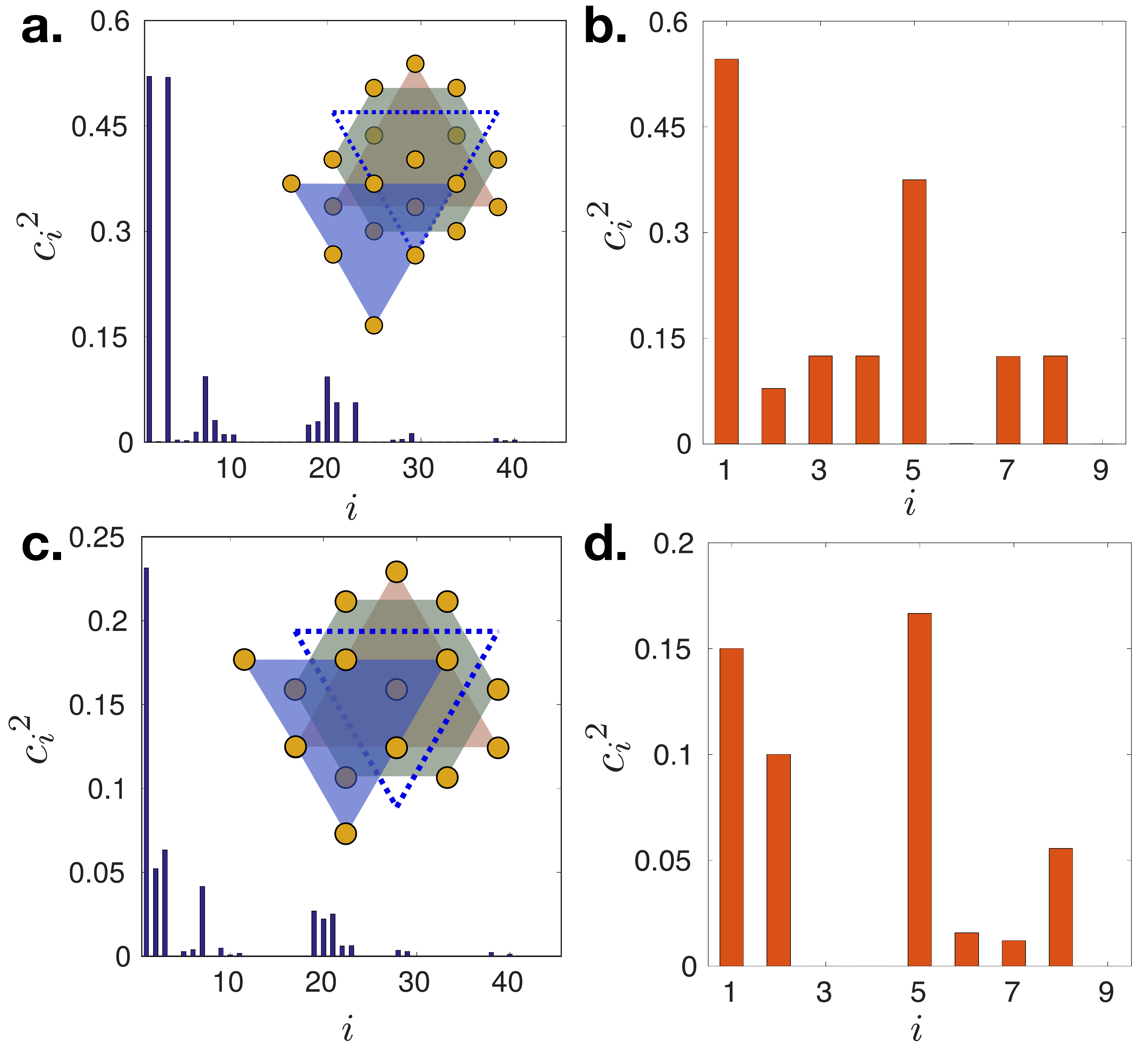}
\caption{Contribution from different {\bf a.}\ non-affine  and {\bf b.}\ affine  modes when one introduces a slip in the FCC lattice. The inset shows a schematic diagram of the lattice as seen from the $[111]$ direction. The original position of the close packed lattice plane is shown as a triangle with a blue dotted boundary and the final position as a blue shaded triangle. Contribution from a stacking fault from {\bf c.} non-affine and {\bf d.} affine modes. The inset shows, as before, the original and final positions of a close packed plane of atoms.}
\label{fig17}
\end{figure}

For the FCC lattice,  we first create a slip along one of the close-packed planes, similar to the triangular case. Fig \ref{fig17} shows the original position (triangle with a dotted boundary) and the new position (blue shaded triangle) of the closed packed plane. The bar-graph of ${c_i}^2$ corresponding to non-affine and affine modes makes it clear that slip in the FCC lattice behaves similarly to a slip in the triangular lattice: we also observe here that the total contributions of the affine and non-affine modes are equal. In the 3d FCC lattice apart from a slip, one can also consider a stacking fault. Fig \ref{fig17} also shows this deformation where the closed packed plane is displaced by half a lattice parameter. The shaded blue region in Fig \ref{fig17} represents the new position of the closed packed plane. Again during a stacking fault it is observed that the non-affine and affine parts contribute equally and the maximum contribution comes from the first three non-affine modes.

This simple exercise therefore strengthens our claim that the non-affine modes that belong to the largest eigenvalues of $\mathsf{PCP}$ are related to fluctuations that tend to nucleate defects. We have seen in~\cite{pnas} that these fluctuations condense under external strain to cause plastic deformation in a 2d triangular crystal. The computations presented here show that non-affine modes deduced from small harmonic fluctuations are able to describe important processes that occur during large deformations. We hope that this knowledge will enable us to study in detail mechanical properties of 3d cubic lattices. Preliminary calculations are under way and will be published elsewhere. 

We believe that our work brings out an interesting aspect concerning defects and the dynamics of deformation in crystalline solids~\cite{CL,rob,hirth}. Atomic fluctuations that generate defects in close packed crystals are shown to be determined by the non-affine modes with the largest eigenvalue. Representing fluctuations in crystals as consisting of smooth phonons and singular defects therefore amounts to making a ``largest eigenvalue approximation'' and neglecting other non-affine modes that make smaller contributions to the total $\chi$. This approximation lies at the heart of all dislocation based theories of crystal plasticity~\cite{rob,hirth,plast1,plast2}. Such an approximation is excellent when the defect-like mode is separated from the others by a large gap in the spectrum of non-affine eigenvalues as in the triangular (Section~\ref{tri}) and FCC (Section~\ref{fcc}) lattices. However, the approach may not work for crystal lattices where such a gap does not exist or is too small, e.g.\ for the planar honeycomb (Section~\ref{phc}) or kagome (Section~\ref{kag}) structures. In amorphous matter also this approximation may not be so useful, even if dislocation-like structures are identifiable~\cite{acharya}. In such cases, new continuum theories of deformation that include {\it all} non-affine modes (or at least a large class of them) may be needed. Such a theory does not exist at present and we hope that our work provides sufficient motivation to the community for thinking along these lines.

\acknowledgements
We acknowledge useful discussions with Saswati Ganguly, Parswa Nath, Madan Rao, Srikanth Sastry, Itamar Procaccia and J\"urgen Horbach.

\newpage
\appendix
\section{Dynamical Matrices}
\label{dynmat}
We give below the expressions for the dynamical matrices for all the lattices studied in this paper. For each case we have included nearest neighbor and next nearest neighbor bonds, with spring constants $k_1$ and $k_2$, respectively. For the triangular lattice only nearest neighbor bonds were retained ($k_2=0$). Separate expressions are given for the additional terms that result in the triangular and honeycomb lattices from the introduction of additional bond-bending terms.
\noindent\\
\subsection{Square} 
\vspace{-1cm}
\begin{flalign}
\cr A_{11}&=2 (k_{1}+k_{2}-\cos (aq_{x}) (k_{1}+k_{2} \cos (aq_{y}))) \nonumber \\
\cr A_{12}&=2 k_{2} \sin (aq_{x}) \sin (aq_{y})=A_{21} \nonumber \\
\cr A_{22}&=2 (k_{1}+k_{2}-\cos (aq_{y}) (k_{1}+k_{2} \cos (aq_{x})))\nonumber \\
\cr D(\mathbf{q})&= \left(
\begin{array}{cc}
 A_{11} & A_{12} \\
 A_{21} & A_{22} \\
\end{array}
\right)
\end{flalign}
\noindent\\
\subsection{Triangular}
\vspace{-0.5cm}
\begin{flalign}
 B_{11}&=k_{1}\bigg(3-2 \cos (aq_{x})-\cos \left(\frac{aq_{x}}{2}\right) \cos \left(\frac{\sqrt{3} aq_{y}}{2}\right)\bigg) \nonumber \\
 B_{12}&=k_{1}\bigg(\sqrt{3} \sin \left(\frac{aq_{x}}{2}\right) \sin \left(\frac{\sqrt{3} aq_{y}}{2}\right)\bigg)=B_{21} \nonumber \\
 B_{22}&=k_{1}\bigg(3-3 \cos \left(\frac{aq_{x}}{2}\right) \cos \left(\frac{\sqrt{3} aq_{y}}{2}\right)\bigg) \nonumber \\
 D(\mathbf{q})&= \left(
\begin{array}{cc}
 B_{11} & B_{12} \\
 B_{21} & B_{22} \\
\end{array}
\right)
\end{flalign}
\ \\
\subsection{Triangular with bond bending:}

\begin{flalign}
B^{\prime}_{11}&=3 k_{b} \bigg[3-\cos (a q_{x}) -2 \cos \left(\frac{a q_{x}}{2}\right) \cos \left(\frac{\sqrt{3} a q_{y}}{2}\right)\bigg] \nonumber \\
&= B^{\prime}_{22} \nonumber \\ \nonumber\\
B^{\prime}_{12}&=0=B^{\prime}_{21} \nonumber \\ \nonumber \\
D(\mathbf{q})&=\left(
\begin{array}{cc}
B_{11} & B_{12} \\
B_{21} & B_{22} \\
\end{array}
\right)
\end{flalign} 
\begin{flalign}
\nonumber
\end{flalign}
\begin{flalign}
\nonumber
\end{flalign}
\vspace{-1cm}
\subsection{Planar honeycomb} 
\begin{flalign}
W_{1}&=\frac{3 k_{1}}{2}+3 k_{2}-3 k_{2} \cos \left(\frac{3 a q_{x}}{2}\right) \cos \left(\frac{\sqrt{3} a q_{y}}{2}\right) \nonumber \\ \nonumber \\
W_{2}&=\sqrt{3} k_{2} \sin \left(\frac{3 a q_{x}}{2}\right) \sin \left(\frac{\sqrt{3} a q_{y}}{2}\right)\nonumber \\ \nonumber \\
W_{3}&=k_{1} \left(-\exp{ (i a q_{x})}-\frac{\exp{\left(-\frac{i a q_{x}}{2}\right)} \cos \left(\frac{\sqrt{3} a q_{y}}{2}\right)}{2}\right)\nonumber \\ \nonumber \\
W_{4}&=\frac{k_{1}}{2} \left(\sqrt{3} i \exp{ \left(-\frac{i a q_{x}}{2}\right)} \sin \left(\frac{\sqrt{3} a q_{y}}{2}\right)\right) \nonumber \\ \nonumber \\
W_{5}&=\frac{3 k_{1}}{2}+3 k_{2}-2 k_{2} \cos \left(\sqrt{3} a q_{y}\right) \nonumber \\
&-k_{2} \cos \left(\frac{3 a q_{x}}{2}\right) \cos \left(\frac{\sqrt{3} a q_{y}}{2}\right)\nonumber \\ \nonumber \\
W_{6}&=\frac{k_{1}}{2}\left(-3 \exp{ \left(-\frac{i a q_{x}}{2}\right)} \cos \left(\frac{\sqrt{3} a q_{y}}{2}\right)\right)\nonumber \\ \nonumber \\
D(\mathbf{q})&=\left(
\begin{array}{cccc}
W_{1} & W_{2} & W_{3} & W_{4} \\
W^{*}_{2} & W_{5} & W_{4} & W_{6} \\
W^{*}_{3} & W^{*}_{4} & W_{1} & W_{2} \\
W^{*}_{4} & W^{*}_{6} & W^{*}_{2} & W_{5}
\end{array}
\right) \\ \nonumber
\end{flalign}
\begin{flalign}
\nonumber
\end{flalign}
\vspace{-10.5cm}
\subsection{Honeycomb with bond bending:}
\vspace{-4cm}
\begin{flalign}
\cr W^{\prime}_{1}&= 3k_{b} \nonumber \\ \nonumber \\
W^{\prime}_{2}&= i \sqrt{3} k_{b} \exp{\left(-\frac{3 i q_{x}a}{2}\right)} \sin \left(\frac{\sqrt{3} q_{y}a}{2}\right)\nonumber \\ \nonumber \\
W^{\prime}_{3}&= -3 k_{b} \exp{\left(-\frac{i q_{x}a}{2}\right)} \cos \left(\frac{\sqrt{3} q_{y}a}{2}\right) \nonumber \\ \nonumber \\
W^{\prime}_{4}&= -2 i \sqrt{3} k_{b} \exp{\left(-\frac{i q_{x}a}{2}\right)} \sin \left(\frac{\sqrt{3} q_{y}a}{2}\right) \nonumber \\ \nonumber \\
W^{\prime}_{5}&= k_{b} \left(7+2 \cos \left(\frac{3 q_{x}a}{2}\right) \cos \left(\frac{\sqrt{3} q_{y}a}{2}\right)\right) \nonumber \\ \nonumber \\
W^{\prime}_{6}&= -3 k_{b} \exp{\left(-\frac{i q_{x}a}{2}\right)} \left(2 e^{\frac{3 i q_{x}a}{2}}+\cos \left(\frac{\sqrt{3} q_{y}a}{2}\right)\right) \nonumber \\ \nonumber \\
W^{\prime}_{7}&= -3 k_{b} \Bigg( 2 \cos (q_{x}a)+\exp{\left(\frac{i q_{x}a}{2}\right)} \cos \left(\frac{\sqrt{3} q_{y}a}{2}\right) \nonumber \\
&  -2 i \sin (q_{x}a) \Bigg) \nonumber \\ \nonumber \\
D(\mathbf{q})&= \left(
\begin{array}{cccc}
W^{\prime}_{1} & W^{\prime}_{2} & W^{\prime}_{3} & W^{\prime}_{4}\\
W^{\prime *}_{2} & W^{\prime}_{5} & W^{\prime}_{4} & W^{\prime}_{6}\\
W^{\prime *}_{3} & W^{\prime *}_{4} & W^{\prime}_{1} & W^{\prime *}_{2}\\
W^{\prime *}_{4} & W^{\prime}_{7} & W^{\prime}_{2} & W^{\prime}_{5}\\
\end{array}
\right)
\end{flalign}
\begin{flalign}
\nonumber
\end{flalign}
\vspace{-1cm}
\subsection{Kagome} 
\begin{flalign}
M_{1}&=k_{1}+3 k_{2} \nonumber \\ \nonumber \\
M_{2}&=0 \nonumber \\ \nonumber \\
M_{3}&=-\frac{1}{4} k_{1} \exp\bigg({-i \bigl(\frac{a q_{x}}{4}+\frac{\sqrt{3}a q_{y}}{4}\bigl)}\bigg) \nonumber \\
&-\frac{1}{4} k_{1} \exp\bigg({-i \bigl(-\frac{a q_{x}}{4}-\frac{\sqrt{3}a q_{y}}{4}\bigl)}\bigg) \nonumber \\
&-\frac{3}{4} k_{2} \exp\bigg({-i \bigl(-\frac{3a q_{x}}{4}+\frac{\sqrt{3}a q_{y}}{4}\bigl)}\bigg) \nonumber \\
&-\frac{3}{4} k_{2} \exp\bigg({-i \bigl(\frac{3a q_{x}}{4}-\frac{\sqrt{3}a q_{y}}{4}\bigl)}\bigg) \nonumber \\ \nonumber \\
M_{4}&=-\frac{1}{4} \sqrt{3} k_{1} \exp\bigg({-i \bigl(\frac{a q_{x}}{4}+\frac{\sqrt{3}a q_{y}}{4}\bigl)}\bigg) \nonumber \\
&-\frac{1}{4} \sqrt{3} k_{1} \exp\bigg({-i \bigl(-\frac{a q_{x}}{4}-\frac{\sqrt{3}a q_{y}}{4}\bigl)}\bigg) \nonumber \\
&+\frac{1}{4} \sqrt{3} k_{2} \exp\bigg({-i \bigl(-\frac{3a q_{x}}{4}+\frac{\sqrt{3}a q_{y}}{4}\bigl)}\bigg) \nonumber \\
&+\frac{1}{4} \sqrt{3} k_{2}\exp\bigg({-i \bigl(\frac{3a q_{x}}{4}-\frac{\sqrt{3}a q_{y}}{4}\bigl)}\bigg) \nonumber \\ \nonumber \\
M_{5}&=-\frac{1}{4} k_{1} \exp\bigg({-i \bigl(-\frac{a q_{x}}{4}+\frac{\sqrt{3}a q_{y}}{4}\bigl)}\bigg) \nonumber \\
&-\frac{1}{4} k_{1} \exp\bigg({-i \bigl(\frac{a q_{x}}{4}-\frac{\sqrt{3}a q_{y}}{4}\bigl)}\bigg) \nonumber \\
&-\frac{3}{4} k_{2} \exp\bigg({-i \bigl(\frac{3a q_{x}}{4}+\frac{\sqrt{3}a q_{y}}{4}\bigl)}\bigg) \nonumber \\
&-\frac{3}{4} k_{2} \exp\bigg({-i \bigl(-\frac{3a q_{x}}{4}-\frac{\sqrt{3}a q_{y}}{4}\bigl)}\bigg) \nonumber \\ \nonumber \\
M_{6}&=\frac{1}{4} \sqrt{3} k_{1} \exp\bigg({-i \bigl(-\frac{a q_{x}}{4}+\frac{\sqrt{3}a q_{y}}{4}\bigl)}\bigg) \nonumber \\
&+\frac{1}{4} \sqrt{3} k_{1} \exp\bigg({-i \bigl(\frac{a q_{x}}{4}-\frac{\sqrt{3}a q_{y}}{4}\bigl)}\bigg) \nonumber \\
&-\frac{1}{4} \sqrt{3} k_{2} \exp\bigg({-i \bigl(\frac{3a q_{x}}{4}+\frac{\sqrt{3}a q_{y}}{4}\bigl)}\bigg) \nonumber \\
&-\frac{1}{4} \sqrt{3} k_{2} \exp\bigg({-i \bigl(-\frac{3a q_{x}}{4}-\frac{\sqrt{3}a q_{y}}{4}\bigl)}\bigg) \nonumber \\ \nonumber \\
M_{7}&=3 k_{1}+k_{2} \nonumber \\ \nonumber \\ \nonumber
\end{flalign}
\begin{flalign}
M_{8}&=-\frac{1}{4} 3 k_{1} \exp\bigg({-i \bigl(\frac{a q_{x}}{4}+\frac{\sqrt{3}a q_{y}}{4}\bigl)}\bigg) \nonumber \\
&-\frac{3}{4} k_{1} \exp\bigg({-i \bigl(-\frac{a q_{x}}{4}-\frac{\sqrt{3}a q_{y}}{4}\bigl)}\bigg) \nonumber \\
&-\frac{1}{4} k_{2} \exp\bigg({-i \bigl(-\frac{3a q_{x}}{4}+\frac{\sqrt{3}a q_{y}}{4}\bigl)}\bigg) \nonumber \\
&-\frac{1}{4} k_{2} \exp\bigg({-i \bigl(\frac{3a q_{x}}{4}-\frac{\sqrt{3}a q_{y}}{4}\bigl)}\bigg) \nonumber \\ \nonumber \\
M_{9}&=-\frac{1}{4} 3 k_{1} \exp\bigg({-i \bigl(-\frac{a q_{x}}{4}+\frac{\sqrt{3}a q_{y}}{4}\bigl)}\bigg) \nonumber \\
&-\frac{3}{4} k_{1} \exp\bigg({-i \bigl(\frac{a q_{x}}{4}-\frac{\sqrt{3}a q_{y}}{4}\bigl)}\bigg) \nonumber \\
&-\frac{1}{4} k_{2} \exp\bigg({-i \bigl(\frac{3a q_{x}}{4}+\frac{\sqrt{3}a q_{y}}{4}\bigl)}\bigg) \nonumber \\
&-\frac{1}{4} k_{2} \exp\bigg({-i \bigl(-\frac{3a q_{x}}{4}-\frac{\sqrt{3}a q_{y}}{4}\bigl)}\bigg) \nonumber \\ \nonumber \\
M_{10}&=\frac{5 k_{1}}{2}+\frac{3 k_{2}}{2} \nonumber \\ \nonumber \\
M_{11}&=\frac{\sqrt{3} k_{1}}{2}-\frac{\sqrt{3} k_{2}}{2} \nonumber \\ \nonumber \\
M_{12}&=-k_{1} \exp\bigl({i \frac{a q_{x}}{2}}\bigl)-k_{1} \exp\bigl({-i \frac{a q_{x}}{2}}\bigl) \nonumber \\ \nonumber \\
M_{13}&=\frac{3 k_{1}}{2}+\frac{5 k_{2}}{2} \nonumber \\ \nonumber \\
M_{14}&=-k_{2} \exp\bigl({i \frac{\sqrt{3}a q_{y}}{2}}\bigl)-k_{2} \exp\bigl({-i \frac{\sqrt{3}a q_{y}}{2}}\bigl) \nonumber \\ \nonumber \\
M_{15}&=\frac{\sqrt{3} k_{2}}{2}-\frac{\sqrt{3} k_{1}}{2} \nonumber \\ \nonumber \\
D(\mathbf{q})&=\left(
\begin{array}{cccccc}
M_{1} & M_{2} & M_{3} & M_{4} & M_{5} & M_{6} \\
M^*_{2} & M_{7} & M_{4} & M_{8} & M_{6} & M_{9} \\
M^*_{3} & M^*_{4} & M_{10} & M_{11} & M_{12} & M_{2} \\
M^*_{4} & M^*_{8} & M^*_{11} & M_{13} & M_{2} & M_{14} \\
M^*_{5} & M^*_{6} & M^*_{12} & M^*_{2} & M_{10} & M_{15} \\
M^*_{6} & M^*_{9} & M^*_{2} & M^*_{14} & M^*_{15} & M_{13} \\
\end{array}
\right)
\end{flalign} 
\subsection{SC}
\vspace{-28mm}
\begin{flalign}
S_{11}&=-2 \cos \left(aq_{x}\right) \left(k_{1}+k_{2} \left( \cos \left(aq_{y}\right) + \cos \left(aq_{z}\right) \right) \right) \nonumber \\
&+ 2k_{1}+4k_{2}\nonumber \\ \nonumber \\
S_{12}&=2 k_{2} \sin \left(aq_{x}\right) \sin \left(aq_{y}\right)=S_{21}\nonumber \\ \nonumber \\
S_{13}&=2 k_{2} \sin \left(aq_{x}\right) \sin \left(aq_{z}\right)=S_{31}\nonumber \\\nonumber \\
S_{22}&=-2 \cos \left(aq_{y}\right) \left(k_{1}+k_{2} \left( \cos \left(aq_{x}\right) + \cos \left(aq_{z}\right) \right) \right) \nonumber\\
&+ 2k_{1}+4k_{2} \nonumber \\ \nonumber \\
S_{23}&=2 k_{2} \sin \left(aq_{y}\right) \sin \left(aq_{z}\right)=S_{32}\nonumber \\ \nonumber \\
S_{33}&=-2 \cos \left(aq_{z}\right) \left(k_{1}+k_{2} \left( \cos \left(aq_{x}\right) + \cos \left(aq_{y}\right) \right) \right) \nonumber \\
&+ 2k_{1}+4k_{2}\nonumber \\ \nonumber \\
D(\mathbf{q})&= \left(
\begin{array}{ccc}
S_{11} & S_{12} & S_{13}\\
S_{21} & S_{22} & S_{23}\\
S_{31} & S_{32} & S_{33}
\end{array}
\right)
\end{flalign}
\begin{flalign}
\nonumber \\ \nonumber
\end{flalign}
\begin{flalign}
\nonumber
\end{flalign}
\vspace{-2cm}\ \\
\subsection{BCC:} 
\begin{flalign}
\cr G_{11}&=-\frac{8}{3}k_{1} \cos \left( \frac{aq_{x}}{2}\right) \cos \left( \frac{aq_{y}}{2}\right) \cos \left( \frac{aq_{z}}{2}\right) \nonumber \\
&-2k_{2} \cos \left(aq_{x}\right)+\frac{8}{3} k_{1} + 2k_{2} \nonumber \\ \nonumber \\
G_{12}&=\frac{8}{3}k_{1} \cos \left( \frac{aq_{z}}{2}\right) \sin \left( \frac{aq_{x}}{2}\right) \sin \left( \frac{aq_{y}}{2}\right)=G_{21}\nonumber \\ \nonumber \\
G_{13}&=\frac{8}{3}k_{1} \cos \left( \frac{aq_{y}}{2}\right) \sin \left( \frac{aq_{x}}{2}\right) \sin \left( \frac{aq_{z}}{2}\right)=G_{31}\nonumber \\ \nonumber \\
G_{22}&=-\frac{8}{3}k_{1} \cos \left( \frac{aq_{x}}{2}\right) \cos \left( \frac{aq_{y}}{2}\right) \cos \left( \frac{aq_{z}}{2}\right) \nonumber \\
&-2k_{2} \cos \left(aq_{y}\right)+\frac{8}{3} k_{1} + 2k_{2}\nonumber \\ \nonumber \\
G_{23}&=\frac{8}{3}k_{1} \cos \left( \frac{aq_{x}}{2}\right) \sin \left( \frac{aq_{y}}{2}\right) \sin \left( \frac{aq_{z}}{2}\right)=G_{32}\nonumber \\ \nonumber \\
G_{33}&=-\frac{8}{3}k_{1} \cos \left( \frac{aq_{x}}{2}\right) \cos \left( \frac{aq_{y}}{2}\right) \cos \left( \frac{aq_{z}}{2}\right) \nonumber \\
&-2k_{2} \cos \left(aq_{z}\right)+\frac{8}{3} k_{1} + 2k_{2}\nonumber \\ \nonumber \\
D(\mathbf{q})&= \left(
\begin{array}{ccc}
G_{11} & G_{12} & G_{13}\\
G_{21} & G_{22} & G_{23}\\
G_{31} & G_{32} & G_{33}\\
\end{array}
\right)
\end{flalign}
\subsection{FCC}
\begin{align}
F_{11}&=4k_{1}+2k_{2}-2k_{2}\cos\left(aq_{x}\right) \nonumber \\
&-2 k_{1} \cos \left(\frac{aq_{x}}{2}\right) \left( \cos \left(\frac{aq_{y}}{2}\right) + \cos \left(\frac{aq_{z}}{2}\right) \right)\nonumber \\ \nonumber \\
 F_{12}&=2 k_{1} \sin \left(\frac{aq_{x}}{2}\right) \sin \left(\frac{aq_{y}}{2}\right)=F_{21}\nonumber \\ \nonumber \\
F_{13}&=2 k_{1} \sin \left(\frac{aq_{x}}{2}\right) \sin \left(\frac{aq_{z}}{2}\right)=F_{31}\nonumber \\ \nonumber \\
F_{22}&=4k_{1}+2k_{2}-2k_{2}\cos\left(aq_{y}\right) \nonumber \\
&-2 k_{1} \cos \left(\frac{aq_{y}}{2}\right) \left( \cos \left(\frac{aq_{x}}{2}\right) + \cos \left(\frac{aq_{z}}{2}\right) \right)\nonumber \\ \nonumber \\
F_{23}&=2 k_{1} \sin \left(\frac{aq_{y}}{2}\right) \sin \left(\frac{aq_{z}}{2}\right)=F_{32}\nonumber \\ \nonumber \\
F_{33}&=4k_{1}+2k_{2}-2k_{2}\cos\left(aq_{z}\right) \nonumber \\
&-2 k_{1} \cos \left(\frac{aq_{z}}{2}\right) \left( \cos \left(\frac{aq_{x}}{2}\right) + \cos \left(\frac{aq_{y}}{2}\right) \right)\nonumber \\ \nonumber \\
D(\mathbf{q})&=\left(
\begin{array}{ccc}
F_{11} & F_{12} & F_{13}\\
F_{21} & F_{22} & F_{23}\\
F_{31} & F_{32} & F_{33}\\
\end{array}
\right)
\end{align}
\vskip 0.5cm

\onecolumngrid
\section{Strain correlation iso-surfaces for 3d lattices}
\label{strain3d}
We give below the equations for the iso-surfaces of the strain-strain correlation functions for the 3d lattices as shown in Fig.~\ref{fig14}. We have used the same notation as in Section~\ref{corr}. 

\subsection{SC}
The iso-surfaces are given by the equations $\beta\langle e_v^2\rangle ({\bf q}) = Q_v^{SC}/Q^{SC} = 1$, $\beta\langle e_u^2\rangle({\bf q}) = Q_u^{SC}/Q^{SC} = 1.5$ and $\beta\langle e_s^2\rangle({\bf q}) = Q_s^{SC}/Q^{SC} = 3.6$, where, 
\begin{subequations}
\begin{flalign}
Q^{SC} & =  a^2 \biggl[k_{1}^3 q_{x}^2 q_{y}^2 q_{z}^2 + k_{1}^2 k_{2} \bigg(q_{x}^4 \bigl(q_{y}^2+q_{z}^2\bigl)+q_{x}^2 \bigl(q_{y}^4 +6 q_{y}^2 q_{z}^2 
+q_{z}^4\bigl) +q_{y}^2 q_{z}^2 \bigl(q_{y}^2+q_{z}^2\bigl)\bigg)+k_{1} k_{2}^2 \bigg(q_{x}^6+5 q_{x}^4 \bigl(q_{y}^2+q_{z}^2\bigl) &\nonumber \\
&+q_{x}^2 \bigl(5 q_{y}^4+3 q_{y}^2 q_{z}^2+5 q_{z}^4\bigl)+q_{y}^6+5 q_{y}^4 q_{z}^2+5 q_{y}^2 q_{z}^4+q_{z}^6\bigg) 
+k_{2}^3 \bigg(2 q_{x}^6+3 q_{x}^4 \bigl(q_{y}^2+q_{z}^2\bigl)+q_{x}^2 \bigl(3 q_{y}^4+8 q_{y}^2 q_{z}^2+3 q_{z}^4\bigl) \nonumber \\ 
&+2 q_{y}^6+3 q_{y}^4 q_{z}^2+3 q_{y}^2 q_{z}^4+2 q_{z}^6\bigg)\bigg]  
\end{flalign}
\begin{flalign}
Q_v^{SC} &=  3 k_{1}^2 q_{x}^2 q_{y}^2 q_{z}^2+2 k_{1} k_{2} \bigg(q_{x}^4 \bigl(q_{y}^2+q_{z}^2\bigl)+q_{x}^2 \bigl(q_{y}^4+q_{z}^4\bigl)
+q_{y}^2 q_{z}^2 \bigl(q_{y}^2+q_{z}^2\bigl)\bigg)+k_{2}^2 \bigg(q_{x}^6+q_{x}^4 \bigl(q_{y}^2+q_{z}^2\bigl)&\nonumber \\
&+q_{x}^2 \bigl(q_{y}^4+3 q_{y}^2 q_{z}^2+q_{z}^4\bigl)+q_{y}^6+q_{y}^4 q_{z}^2+q_{y}^2 q_{z}^4+q_{z}^6\bigg) 
\end{flalign}
\begin{flalign}
Q_u^{SC}  &=  3 k_{1}^2 q_{x}^2 q_{y}^2 q_{z}^2+2 k_{1} k_{2} \bigg(q_{x}^4 \bigl(q_{y}^2+q_{z}^2\bigl)+q_{x}^2 \bigl(q_{y}^4+8 q_{y}^2 q_{z}^2 
+q_{z}^4\bigl)+q_{y}^2 q_{z}^2 \bigl(q_{y}^2+q_{z}^2\bigl)\bigg)+k_{2}^2 \bigg(q_{x}^6+9 q_{x}^4 \bigl(q_{y}^2+q_{z}^2\bigl) & \nonumber \\
&+3 q_{x}^2\bigl(3 q_{y}^4+q_{y}^2 q_{z}^2+3 q_{z}^4\bigl)+q_{y}^6+q_{y}^4 q_{z}^2+q_{y}^2 q_{z}^4+q_{z}^6\bigg)
\end{flalign}
\begin{flalign}
Q_s^{SC} &=  k_{1}^2 q_{z}^2 \bigl(q_{x}^4+q_{y}^4\bigl)+k_{1} k_{2} \bigg(q_{x}^6+q_{x}^4 \bigl(q_{y}^2+4 q_{z}^2\bigl)
+q_{x}^2 \bigl(q_{y}^2-q_{z}^2\bigl)^2+q_{y}^2 \bigl(q_{y}^4+4 q_{y}^2 q_{z}^2+q_{z}^4\bigl)\bigg) 
+k_{2}^2 \bigg(2 q_{x}^6+q_{x}^4 q_{z}^2 
&\nonumber \\
&+2 q_{x}^2 \bigl(3 q_{y}^2 q_{z}^2+q_{z}^4\bigl)+2 q_{y}^6
+q_{y}^4 q_{z}^2+2 q_{y}^2 q_{z}^4\bigg) 
\end{flalign}
\end{subequations}

\subsection{BCC}
The iso-surfaces are given by the equations $\beta\langle e_v^2\rangle ({\bf q}) = Q_v^{BCC}/Q^{BCC} = 1.6$, $\beta\langle e_u^2\rangle({\bf q}) = Q_u^{BCC}/Q^{BCC} = 5.5$ and $\beta\langle e_s^2\rangle({\bf q}) = Q_s^{BCC}/Q^{BCC}=2.15$, where, 
\begin{subequations}
\begin{flalign}
Q^{BCC}&=a^2 \bigg[k_{1}^3 \bigg(-q_{z}^4 \bigl(q_{x}^2+q_{y}^2\bigl)+\bigl(q_{x}^2-q_{y}^2\bigl)^2 \bigl(q_{x}^2+q_{y}^2\bigl)-q_{z}^2 \bigl(q_{x}^4 
-10 q_{x}^2 q_{y}^2+q_{y}^4\bigl)+q_{z}^6\bigg)+3 k_{1}^2 k_{2} \bigg(q_{x}^6+3 q_{x}^4 \bigl(q_{y}^2+q_{z}^2\bigl) &\nonumber \\
&+3 q_{x}^2 \bigl(q_{y}^2-q_{z}^2\bigl)^2+\bigl(q_{y}^2+q_{z}^2\bigl)^3\bigg)+9 k_{1} k_{2}^2 \bigl(q_{x}^2 +q_{y}^2
+q_{z}^2\bigl)\bigg(q_{x}^2 \bigl(q_{y}^2+q_{z}^2\bigl)+q_{y}^2 q_{z}^2\bigg)+27 k_{2}^3 q_{x}^2 q_{y}^2 q_{z}^2\bigg]
\end{flalign}
\begin{flalign}
Q_v^{BCC}&=3 k_{1} \bigl(q_{x}^2+q_{y}^2\bigl) \bigg(6 k_{2} q_{x}^2
q_{y}^2+k_{1} \bigl(q_{x}^2-q_{y}^2\bigl)^2\bigg) 
-3 \bigg(k_{1}
(k_{1}-6 k_{2}) q_{x}^4-3 \bigl(2 k_{1}^2-6 k_{1} k_{2}+9
k_{2}^2\bigl) &\nonumber \\
& q_{x}^2 q_{y}^2+k_{1} (k_{1}-6 k_{2})
q_{y}^4\bigg) q_{z}^2-3 k_{1} (k_{1}-6 k_{2}) 
\bigl(q_{x}^2+q_{y}^2\bigl) q_{z}^4+3 k_{1}^2 q_{z}^6
\end{flalign}
\begin{flalign}
Q_u^{BCC} &=3 \bigg[27 k_{2}^2 q_{x}^2 q_{y}^2 q_{z}^2+6 k_{1} k_{2}
   \bigg(q_{x}^4 \bigl(q_{y}^2+q_{z}^2\bigl)+q_{y}^2 q_{z}^2
   \bigl(q_{y}^2 
+q_{z}^2\bigl)+q_{x}^2 \bigl(q_{y}^4+5 q_{y}^2
   q_{z}^2+q_{z}^4\bigl)\bigg)+k_{1}^2 \bigg(q_{x}^6+7 q_{x}^4
   \bigl(q_{y}^2 +q_{z}^2\bigl)&\nonumber \\
&+\bigl(q_{y}^2-q_{z}^2\bigl)^2
   \bigl(q_{y}^2+q_{z}^2\bigl)+q_{x}^2 \bigl(7 q_{y}^4-10 q_{y}^2
   q_{z}^2 
   +7 q_{z}^4\bigl)\bigg)\bigg]
\end{flalign} 
\begin{flalign}  
Q_s^{BCC}&=3 \bigg[9 k_{2}^2 \bigl(q_{x}^4+q_{y}^4\bigl) q_{z}^2+3 k_{1}
   k_{2} \bigg(q_{x}^6+q_{x}^2
   \bigl(q_{y}^2-q_{z}^2\bigl)^2 
 +q_{y}^2\bigl(q_{y}^2+q_{z}^2\bigl)^2+q_{x}^4 \bigl(q_{y}^2+2
   q_{z}^2\bigl)\bigg)& \nonumber \\
&   +k_{1}^2 \bigg(q_{x}^6 
-q_{x}^4\bigl(q_{y}^2+2 q_{z}^2\bigl)+\bigl(q_{y}^3-q_{y}
q_{z}^2\bigl)^2 
+q_{x}^2 \bigl(-q_{y}^4+8 q_{y}^2
   q_{z}^2+q_{z}^4\bigl)\bigg)\bigg]
\end{flalign}
\end{subequations}
\subsection{FCC}
The iso-surfaces are given by the equations $\beta\langle e_v^2\rangle ({\bf q}) = Q_v^{FCC}/Q^{FCC}=1.44$, $\beta\langle e_u^2\rangle({\bf q}) = Q_u^{FCC}/Q^{FCC}=3.6$ and $\beta\langle e_s^2\rangle({\bf q}) = Q_s^{FCC}/Q^{FCC}=2.75$, where, 
\begin{subequations}
\begin{flalign}
Q^{FCC} &=  a^2 \bigg[k_{1}^3 \bigg(2 q_{x}^6+3 q_{x}^4 \bigl(q_{y}^2+q_{z}^2\bigl)+q_{x}^2 \bigl(3 q_{y}^4+8 q_{y}^2 q_{z}^2+3 q_{z}^4\bigl) 
+2 q_{y}^6+3 q_{y}^4 q_{z}^2+3 q_{y}^2 q_{z}^4+2 q_{z}^6\bigg)+4 k_{1}^2 k_{2} \bigg(q_{x}^6+5 q_{x}^4 \bigl(q_{y}^2 
& \nonumber \\
&+q_{z}^2\bigl)+q_{x}^2 \bigl(5 q_{y}^4+3 q_{y}^2 q_{z}^2+5 q_{z}^4\bigl)+q_{y}^6+5 q_{y}^4 q_{z}^2+5 q_{y}^2 q_{z}^4 
+q_{z}^6\bigg)+16 k_{1} k_{2}^2 \bigg(q_{x}^4 \bigl(q_{y}^2+q_{z}^2\bigl)+q_{x}^2 \bigl(q_{y}^4+6 q_{y}^2 q_{z}^2+q_{z}^4\bigl) \nonumber \\ 
&+q_{y}^2 q_{z}^2 \bigl(q_{y}^2+q_{z}^2\bigl)\bigg)+64 k_{2}^3 q_{x}^2 q_{y}^2 q_{z}^2\bigg] 
\end{flalign}
\begin{flalign}
Q_v^{FCC}  &=  4 \bigg[k_{1}^2 \bigg(q_{x}^6+q_{x}^4 \bigl(q_{y}^2+q_{z}^2\bigl)+q_{x}^2 \bigl(q_{y}^4+3 q_{y}^2 q_{z}^2+q_{z}^4\bigl) 
+q_{y}^6+q_{y}^4 q_{z}^2+q_{y}^2 q_{z}^4+q_{z}^6\bigg)+8 k_{1} k_{2} \bigg(q_{x}^4 \bigl(q_{y}^2+q_{z}^2\bigl) &\nonumber \\
&+q_{x}^2 \bigl(q_{y}^4+q_{z}^4\bigl)+q_{y}^2 q_{z}^2 \bigl(q_{y}^2+q_{z}^2\bigl)\bigg)+48 k_{2}^2 q_{x}^2 q_{y}^2 q_{z}^2\bigg] 
\end{flalign}
\begin{flalign}
Q_u^{FCC}  &=  4 \bigg[k_{1}^2 \bigg(q_{x}^6+9 q_{x}^4 \bigl(q_{y}^2+q_{z}^2\bigl)+3 q_{x}^2 \bigl(3 q_{y}^4+q_{y}^2 q_{z}^2
+3 q_{z}^4\bigl)+q_{y}^6+q_{y}^4 q_{z}^2+q_{y}^2 q_{z}^4+q_{z}^6\bigg)+8 k_{1} k_{2} \bigg( q_{x}^4 \bigl(q_{y}^2+q_{z}^2\bigl) &\nonumber \\
&
+q_{x}^2 \bigl(q_{y}^4+8 q_{y}^2 q_{z}^2+q_{z}^4\bigl)+q_{y}^2 q_{z}^2 \bigl(q_{y}^2 
+q_{z}^2\bigl)\bigg)+48 k_{2}^2 q_{x}^2 q_{y}^2 q_{z}^2\bigg]
\end{flalign} 
\begin{flalign}  
Q_s^{FCC}  &=  4 \bigg[k_{1}^2 \bigg(2 q_{x}^6+q_{x}^4 q_{z}^2+2 q_{x}^2 \bigl(3 q_{y}^2 q_{z}^2+q_{z}^4\bigl)+2 q_{y}^6+ 
q_{y}^4 q_{z}^2+2 q_{y}^2 q_{z}^4\bigg)+4 k_{1} k_{2} \bigg(q_{x}^6+q_{x}^4 \bigl(q_{y}^2+4 q_{z}^2\bigl) &\nonumber \\
 &+q_{x}^2\bigl(q_{y}^2-q_{z}^2\bigl)^2+q_{y}^2 \bigl(q_{y}^4+4 q_{y}^2 q_{z}^2+q_{z}^4\bigl)\bigg) 
 +16 k_{2}^2 q_{z}^2 \bigl(q_{x}^4+q_{y}^4\bigl)\bigg]
\end{flalign}
\end{subequations}

\twocolumngrid
\footnotesize{

}

\end{document}